\documentclass[useAMS,usenatbib,usegraphicx]{mn2e}
\usepackage{amssymb}
\usepackage{times}
\usepackage{aas_macros} 
\usepackage{epsfig}
\usepackage[usenames]{color}

\newcommand*{\mathcolor}{}
\def\mathcolor#1#{\mathcoloraux{#1}}
\newcommand*{\mathcoloraux}[3]{%
  \protect\leavevmode
  \begingroup
    \color#1{#2}#3%
  \endgroup
}

\title[Asteroseismic measurement of rotation in KIC~9244992]{Asteroseismic measurement of slow, nearly-uniform surface-to-core rotation in the main sequence F star  KIC~9244992}

\author[Saio et al.]
{Hideyuki Saio$^{1}$, Donald W. Kurtz$^2$, Masao Takata$^3$, Hiromoto Shibahashi$^3$,   \newauthor{Simon J. Murphy$^{4,5}$, Takashi Sekii$^{6}$, 
Timothy R. Bedding$^{4,5}$} \\
$^{1}$Astronomical Institute, Graduate School of Science, Tohoku University, Sendai, Miyagi 980-8578, Japan \\
$^{2}$Jeremiah Horrocks Institute, University of Central
Lancashire, Preston PR1 2HE, UK\\
$^{3}$Department of Astronomy, School of Science, The University of Tokyo, Bunkyo-ku, Tokyo 113-0033, Japan \\
$^{4}$Sydney Institute for Astronomy, School of Physics, The University of Sydney, NSW 2006, Australia \\
$^{5}$Stellar Astrophysics Centre, Department of Physics and Astronomy, Aarhus
University, Denmark\\
$^{6}$National Astronomical Observatory of Japan, 2-21-1 Osawa, Mitaka, Tokyo 181-8588, Japan
}

\begin{document}

\maketitle

\begin{abstract}
We have found a rotationally split series of core g-mode triplets and surface p-mode multiplets in a main sequence F star, KIC~9244992. Comparison with models shows that  the star has a mass of about 1.45~M$_\odot$, and is at an advanced stage of main sequence evolution  in which the central hydrogen abundance mass fraction is reduced to about 0.1.  This is the second case, following KIC~11145123, of an asteroseismic determination of the rotation of the deep core and surface of an A-F main-sequence star. We have found, essentially model-independently, 
that the rotation near the surface, obtained from p-mode splittings, is 66~d, slightly slower than the rotation of 64~d in the core, measured by g-mode splittings. KIC~9244992 is similar to KIC~11145123 in that both are near the end of  main-sequence stage with very slow and nearly uniform rotation.  This indicates the angular momentum transport in the interior of an A-F star during the main sequence stage is much stronger than that expected from standard theoretical formulations. 
\end{abstract}


\section{Introduction}
\label{sec:1}
  
The internal rotation of stars, which undoubtedly affects the evolution of all stars in various ways, is only poorly understood 
and remains one of the unsolved fundamental problems in stellar astrophysics. The development of asteroseismic investigation is changing the situation,  making it now possible to observe the previously invisible stellar internal rotation. 

Stellar rotation affects the frequency spectrum of stellar oscillation modes. It induces a multiplet fine structure, for which the frequency separations of the multiplet components are dependent on the internal rotation profile of the star, as well as the stellar structure. In principle, the rotation profile can be inferred by carefully studying the oscillation frequency spectrum.

The first success was the case of the Sun, for which a spatially resolved image is observable, hence a large number of p~modes with a wide range of horizontal scales can be detected. It was shown, through the measurement and analysis of frequencies of the solar p~modes, 
that the convective envelope of the Sun rotates differentially with respect to latitude, but with relatively little radial dependence,  while the radiative interior, at least down to $\sim$40~per~cent in radius from the centre, rotates nearly uniformly at a rate slightly slower than the surface equatorial rate \citep{schou1998, Korzennik2012}. Due to lack of firm detection of g~modes, which are sensitive to the much deeper interior, the rotation profile near the solar centre is as yet unknown. Nevertheless, the helioseismic result showing nearly uniform rotation in the bulk of radiative interior was a surprise, since the deep interior was expected to rotate more rapidly than the envelope as a consequence of gradual evolutionary contraction of the core, from a naive consideration assuming local conservation of angular momentum. This finding stimulated theoretical attempts to explain the mechanism of angular momentum transport in the Sun \citep[e.g.,][]{GoughMcIntyre98,Talon2002, Mathis2008}.

The almost uninterrupted continuous photometry with unprecedented low noise level carried out by space missions opened a new window to unveil the internal rotation of distant stars. Although the number of detected oscillation modes in unresolved distant stars is much smaller than the solar case, in many sub-giants and red giants, so-called `mixed' modes have been detected, in addition to p~modes. Those mixed modes have dual characters of p~modes and g~modes and provide us with information of the rotation rate in the deep interior. It was then clearly confirmed that, in several sub-giants and red giants, their cores rotate faster than the envelopes \citep{beck2012,deheuvels2012, deheuvels2014,mosser2012}. However, the contrast between the core and the surface is weaker than expected. Therefore, a strong mechanism for angular momentum transport must be acting at some evolutionary stages before stars become red giants.
It is then important to investigate systematically the interior rotation of stars in the earlier evolutionary stages.

With the aim to observe the interior rotation of main-sequence stars,
our previous study \citep{kurtz14} revealed, for the first time, the rotation of a main sequence A star, KIC~11145123, at the surface and in the core. We found it to be nearly a rigid rotator with a rotation period near to 100~d and with the surface rotating slightly faster than the core. In addition, we found from the period spacings of consecutive radial order of g~modes that  the star is contracting at the evolutionary stage of core hydrogen exhaustion; i.e.,the star is near to the terminal-age main-sequence.  Such remarkable findings of the internal rotation and the evolutionary stage of  KIC~11145123 were possible because the star has a rich frequency spectrum with many rotational multiplets for both p~modes and g~modes. We have now found another Kepler star, KIC~9244992, to have similarly rich frequencies in g~modes as well as p~modes, which makes another detailed analysis possible.  This paper presents the results of our analysis on this star.

\subsection{KIC~9244992}

KIC~9244992 has a {\it Kepler} magnitude ${\rm Kp} = 14$. From the Kepler Input Catalogue (KIC) revised photometry \citep{huberetal2014}, its effective temperature is $T_{\rm eff} = 6900 \pm 300$~K ($\log T_{\rm eff} = 3.839 \pm 0.018$) and its surface gravity is $\log g = 3.5 \pm 0.4$ (cgs units), showing it to be a main-sequence,  or possibly post-main-sequence F star. The metallicity is estimated to be [Fe/H]$=-0.15\pm 0.30$. Based on its high tangential velocity,
 \citet{balona12} included this star in their list of  SX~Phe candidates. A recent spectroscopic study by Nemec et al. (in preparation) gives 
$T_{\rm eff} = 7000 ^{+300}_{-100}$~K, $\log g = 3.8 ^{+0.2}_{-0.3}$, 
 and $v \sin i < 6$~km~s$^{-1}$. 
KIC~9244992 is somewhat cooler than  KIC~11145123 ($T_{\rm eff}\approx 8000$~K). 

\section{Observations and frequency analysis}
\label{sec:obs}

The data used for the analysis in this paper are the {\it Kepler} quarters 1 to 17 (Q1 -- Q17) long cadence (LC) data, which include all of the available data. The star was not observed in Q0, nor was it observed in short cadence. {\it Kepler} has an orbital period about the Sun of 372.4536~d, hence the quarters are just over 93~d. We used the multi-scale, maximum a posteriori (msMAP) pipeline data; information on the reduction pipeline can be found in the data release 21 notes\footnote{https://archive.stsci.edu/kepler/data\_release.html}. To optimise the search for exoplanet transit signals, the msMAP data pipeline removes or alters astrophysical signals with frequencies less than 0.1~d$^{-1}$ (or periods greater than 10~d). A highpass filter was run to remove noise below 0.2~d$^{-1}$, where there is no stellar signal. None of the pulsation frequencies we analyse in this paper are near to that lower limit, but if the star has a direct rotational signal, e.g. from starspots, that will have been erased by the pipeline and our high-pass filter. 

We extracted pulsation frequencies first with a Discrete Fourier Transform amplitude spectrum analysis \citep{kurtz1985}, then by fitting the obtained frequencies by least-squares, then non-linear least-squares, along with all previously identified frequencies at each step. We selected a portion of the amplitude spectrum that showed only white noise to estimate the uncertainties of the frequencies and phases that are listed in tables below. These uncertainties are similar to those that would be obtained by pre-whitening all significant peaks from the amplitude spectrum and then fitting all frequencies by least-squares. They do not include the effects of possible unresolved peaks, which could shift the frequency up to $1.7\times10^{-4}$~d$^{-1}$ for the 4-y length of the data \citep{kallinger08}, in the worst case.

\begin{figure}
\centering
\includegraphics[width=0.9\linewidth,angle=0]{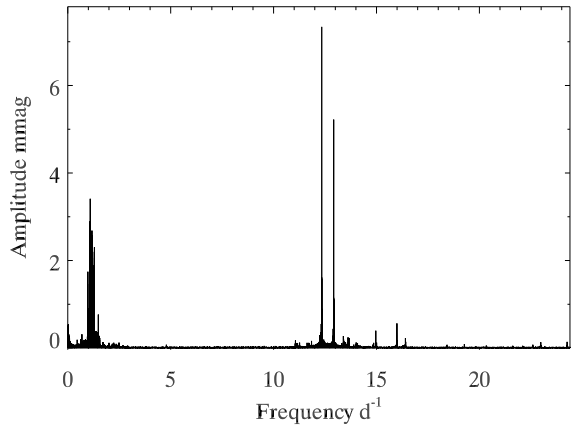}\\ 
\caption{
The amplitude spectrum for the Q1 to Q17 {\it Kepler} long cadence data up to nearly the Nyquist frequency for KIC~9244992, showing the clearly separated p-mode and g-mode frequency range.}
\label{fig:9244992_ft-all}
\end{figure}

Fig.~\ref{fig:9244992_ft-all} shows a full amplitude spectrum nearly out to the Nyquist frequency for KIC~9244992 for the nearly continuous {\it Kepler} Q1-17 LC data spanning 1459~d (4.0~y). There are pulsations in both the g-mode  and p-mode frequency regions, which are clearly separated. In the p-mode frequency range the highest amplitude peak is a singlet, presumably from a radial mode, and there are six triplets and an isolated quintuplet (or septuplet) split by rotation. The g-mode frequency range is dominated by a series of 17 high-overtone dipole mode triplets split by the rotation frequency in the core of the star.

\subsection{The g~modes}
\label{sec:gmodes}

\begin{figure}
\centering
\includegraphics[width=0.9\linewidth,angle=0]{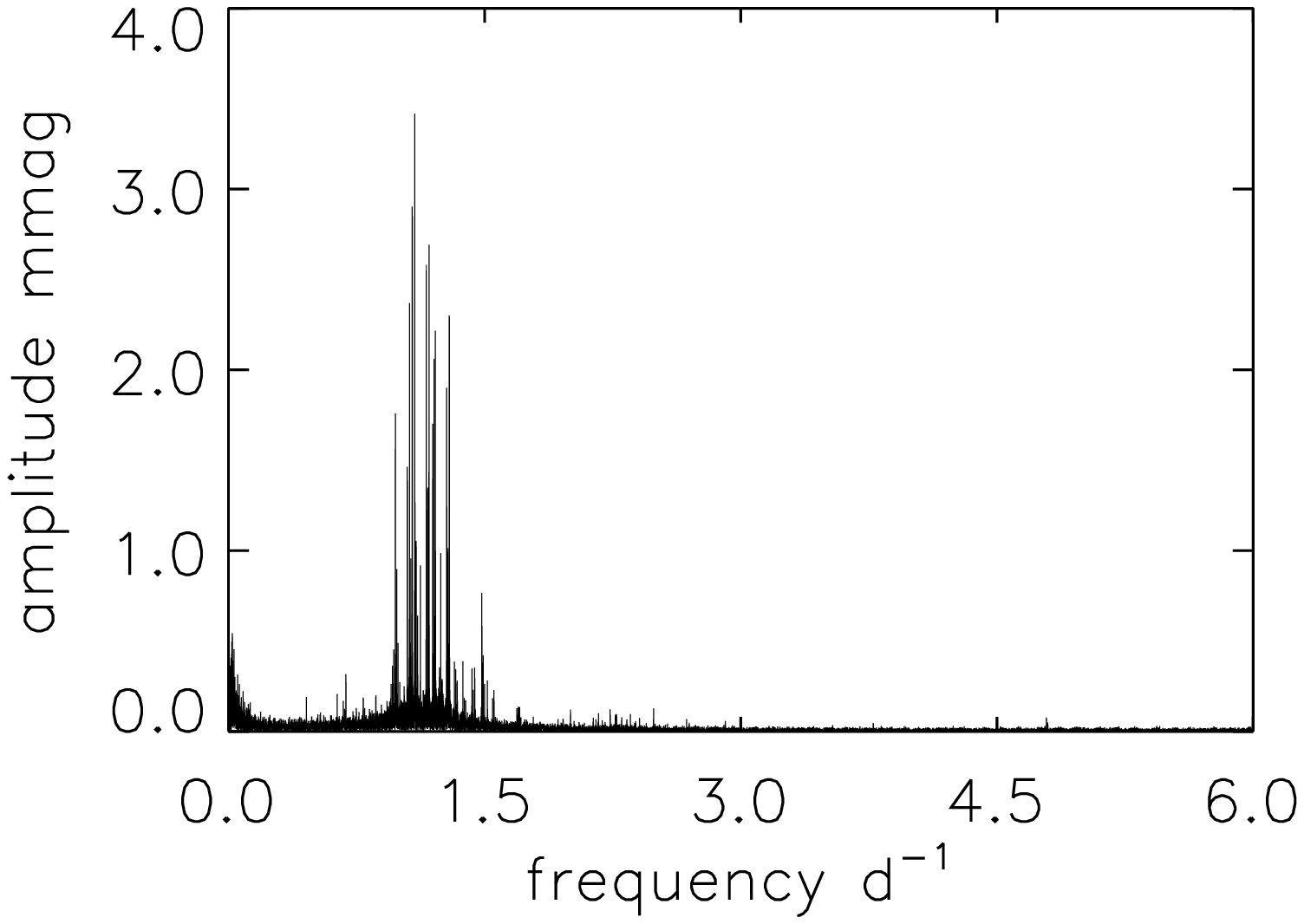}\\ 
\includegraphics[width=0.9\linewidth,angle=0]{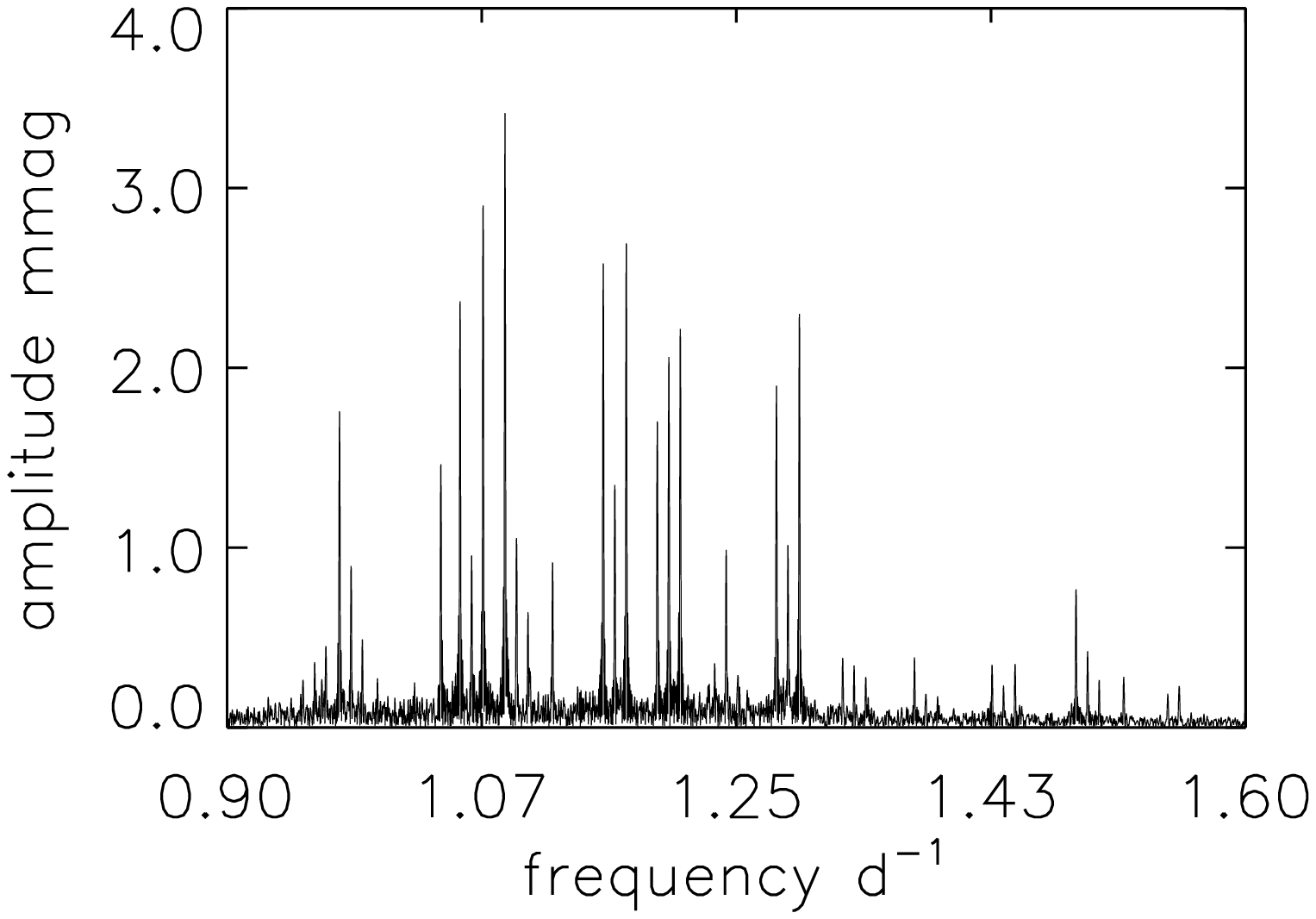}\\ 
\includegraphics[width=0.9\linewidth,angle=0]{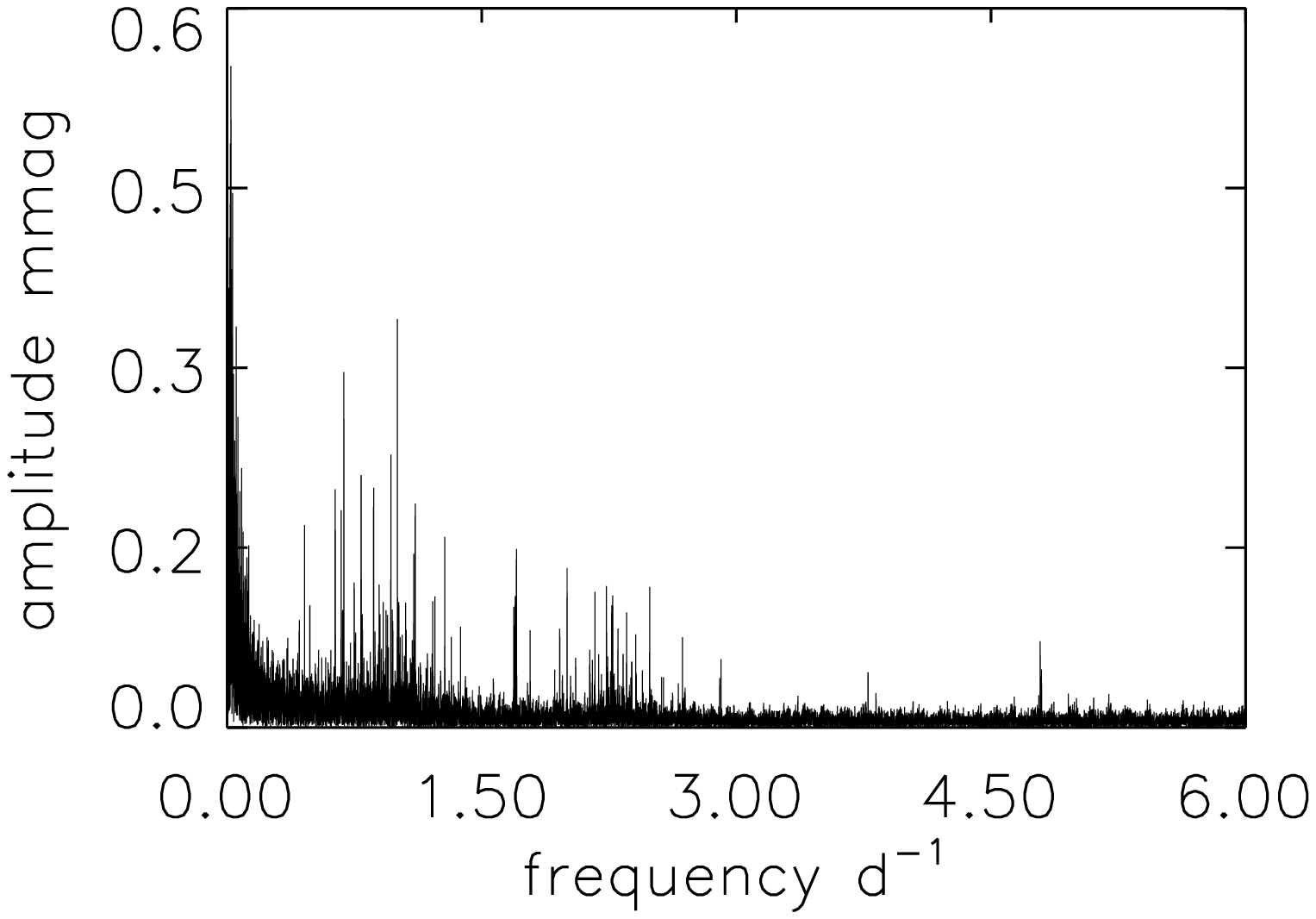}   
\caption{Top panel: An amplitude spectrum for the g-mode frequency range. Middle panel: The range of the higher amplitude g-mode peaks. Many triplets can be seen. From this range we have obtained 14 triplets split nearly equally, and 3 other modes. Bottom panel: An amplitude spectrum of the residuals after pre-whitening by 45 frequencies. Note the change in the vertical scale from the top two panels. An additional 5 triplets are obtained in this range.
}
\label{fig:9244992_ftg}
\end{figure}

\begin{figure}
\centering
\includegraphics[width=0.9\linewidth,angle=0]{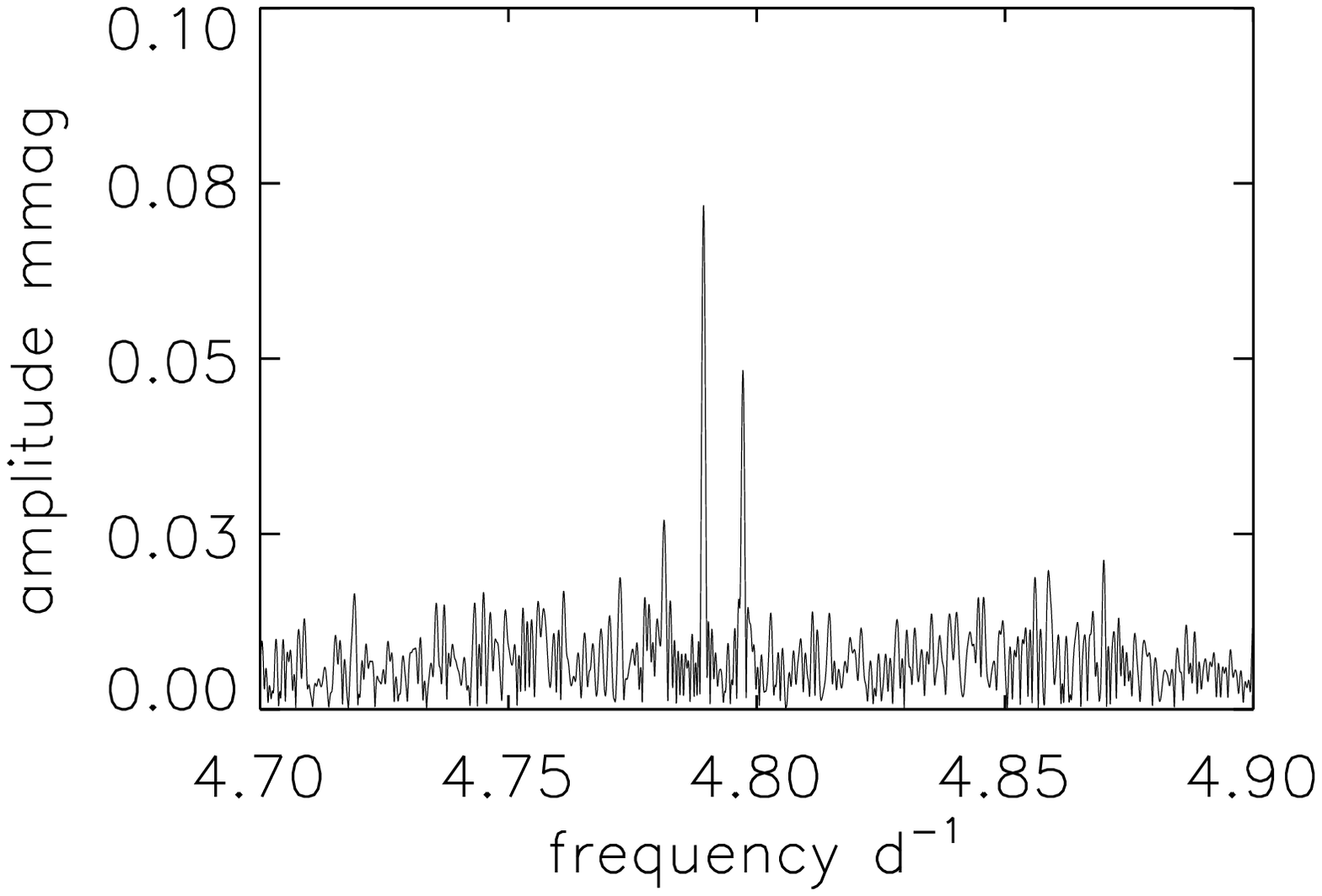}\\ 
\includegraphics[width=0.9\linewidth,angle=0]{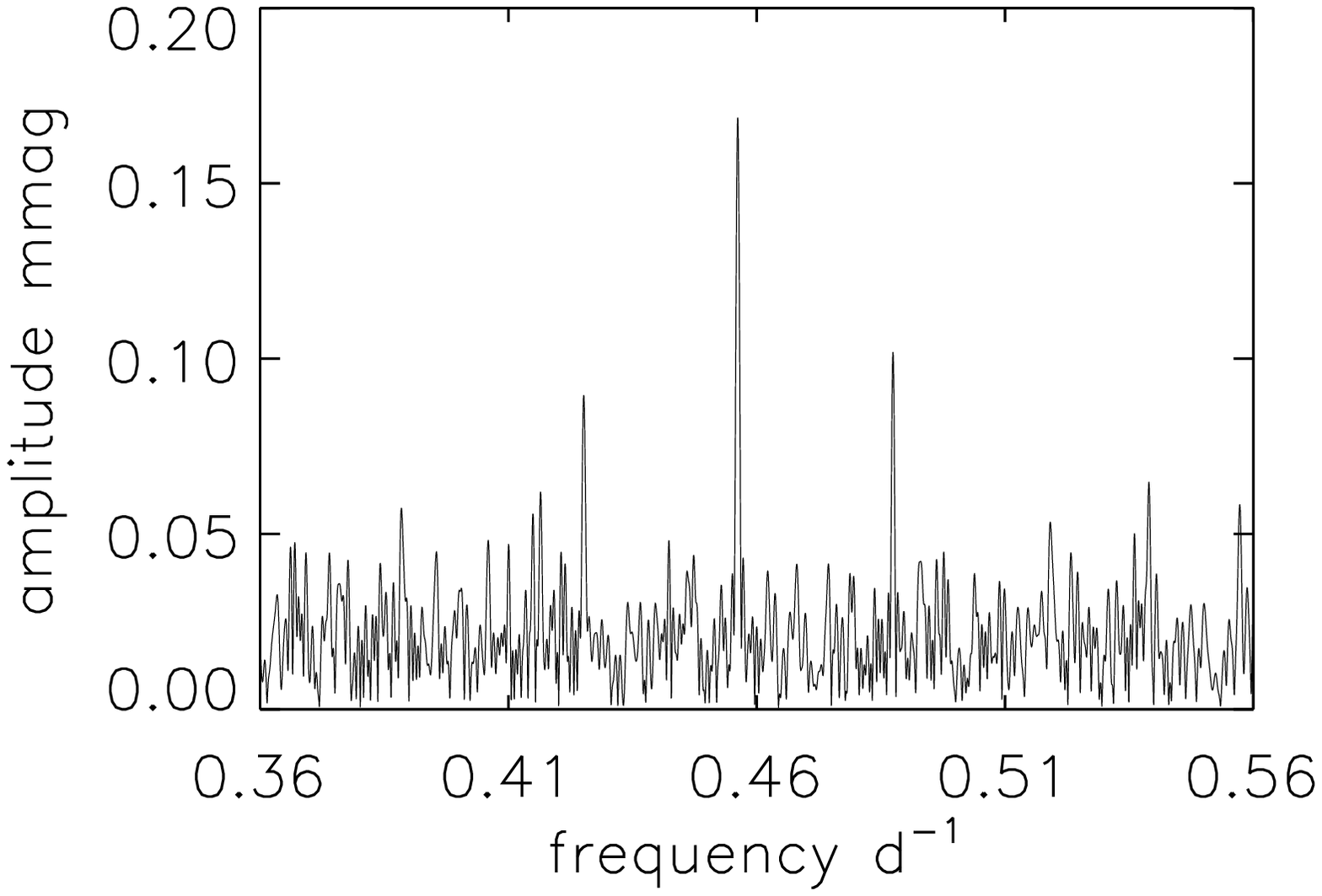}   
\caption{
Top panel: An amplitude spectrum for the $4.8$-d$^{-1}$ triplet.
Bottom panel: A low frequency triplet with a different splitting (see text for details).
}
\label{fig:9244992_ftg2}
\end{figure}

Fig.~\ref{fig:9244992_ftg} shows amplitude spectra for the g-mode frequency range. High-amplitude g~modes lie in the frequency range  $0.9 - 1.6$~d$^{-1}$ (middle panel). A first search for high-amplitude g~modes in this range yielded 45 frequencies: 14 triplets and 3 other frequencies. The bottom panel shows the amplitude spectrum after removing these 45 frequencies. From the pre-whitened data we found  additional five triplets. Among the triplets obtained, two at $0.45$~d$^{-1}$ and $4.8$~d$^{-1}$ (see Fig.~\ref{fig:9244992_ftg2}) are notably far from the main g-mode frequency range. The higher frequency mode, having rotational splittings similar to the high-amplitude g~modes, can be identified as a relatively low order dipole g~mode, while the splitting of the lower frequency triplet is much larger so that the mode identification is unclear. The bottom panel of Fig.~\ref{fig:9244992_ftg} indicates the presence of more pulsation modes in the g-mode range. However, we have stopped searching for further small amplitude modes because of smaller signal-to-noise ratios. 

Tables~\ref{table:9244992_g} and \ref{table:9244992_g_add} give the results of a combination of linear least-squares and non-linear least-squares fits of g-mode frequencies to the Q1-17 data. Because of the decreasing signal-to-noise ratio for these further peaks, we chose for this first study of KIC~9244992 to analyse only the most significant multiplets. The first column in Tables~\ref{table:9244992_g} and \ref{table:9244992_g_add} marks the g~modes (g). The second column gives azimuthal order $m$, where we adopt the convention that $m > 0$ corresponds to a prograde mode. The next three columns give frequency, amplitude and phase (with respect to $t_0 = {\rm BJD}~2455694.25$). The sixth column gives the frequency separations between the components of each multiplet, which correspond to the rotational splitting. The splittings show small asymmetries; i.e., $\nu(m=0)-\nu(m=-1) > \nu(m=1)-\nu(m=0)$ with few exceptions. The asymmetries are up to $\sim 1.6\times 10^{-4}$~d$^{-1}$, and systematically larger for lower frequencies. They are larger  than those expected from the second-order effect of the Coriolis force, and the sign is in the opposite sense. The cause is not clear.

\begin{table*}
\centering
\caption[]{A least-squares fit of  the g-mode frequency multiplets for KIC~9244992. The phases are measured with respect to $t_0 = {\rm BJD}~2455694.25$
 }
\begin{tabular}{crccrc}
\hline
\hline
& $m$ & \multicolumn{1}{c}{frequency} & \multicolumn{1}{c}{amplitude} &
\multicolumn{1}{c}{phase} & \multicolumn{1}{c}{$\Delta \nu_{\rm rot}$} \\
& & \multicolumn{1}{c}{d$^{-1}$} & \multicolumn{1}{c}{mmag} &
\multicolumn{1}{c}{radians} & \multicolumn{1}{c}{d$^{-1}$} \\
\hline
g &$-1$ & $0.9521792 \pm 0.0000092$ & $0.2077 \pm 0.0050$ & $-1.2300 \pm 0.0242$
& \\
g &0 & $0.9601528 \pm 0.0000068$ & $0.2818 \pm 0.0050$ & $-2.5378 \pm 0.0178$ &	  $0.0079736 \pm 0.0000114$ \\
g &1 & $0.9679628 \pm 0.0000040$ & $0.4851 \pm 0.0050$ & $-1.5742 \pm 0.0104$ &	 $0.0078100 \pm 0.0000079$ \\
\hline
g &$-1$& $0.9772590 \pm 0.0000010$ & $1.8347 \pm 0.0050$ & $-1.0530 \pm 0.0027$ &\\
g &0 & $0.9852006 \pm 0.0000021$ & $0.9153 \pm 0.0050$ & $-2.1938 \pm 0.0055$ &	 $0.0079415 \pm 0.0000023$ \\
g &1& $0.9930256 \pm 0.0000038$ & $0.5043 \pm 0.0050$ & $-0.3547 \pm 0.0100$ &	 $0.0078250 \pm 0.0000043$ \\
\hline
g &$-1$ & $1.0034941 \pm 0.0000055$ & $0.3464 \pm 0.0050$ & $2.6150 \pm 0.0145$ & \\
g & 0 & $1.0114562 \pm 0.0000184$ & $0.1041 \pm 0.0050$ & $1.2851 \pm 0.0482$ & $0.0079621 \pm 0.0000192$ \\
g & 1 & $1.0192389 \pm 0.0000238$ & $0.0804 \pm 0.0050$ & $-2.2112 \pm 0.0626$ & $0.0077827 \pm 0.0000301$ \\
\hline
g &1 & $1.0468674 \pm 0.0000013$ & $1.4910 \pm 0.0050$ & $-0.9497 \pm 0.0034$ & \\
\hline
g &$-1$ & $1.0601156 \pm 0.0000008$ & $2.2770 \pm 0.0050$ & $-1.7949 \pm 0.0022$ & \\
g &0 & $1.0680731 \pm 0.0000018$ & $1.0493 \pm 0.0050$ & $3.0187 \pm 0.0048$ & 	 $0.0079574 \pm 0.0000020$ \\
g &1 & $1.0759049 \pm 0.0000007$ & $2.9357 \pm 0.0050$ & $2.8047 \pm 0.0017$ &	$0.0078318 \pm 0.0000019$ \\
\hline
g & $-1$ & $1.0909822 \pm 0.0000006$ & $3.3646 \pm 0.0050$ & $-0.9789 \pm 0.0015$ \\
g & 0 & $1.0988823 \pm 0.0000019$ & $1.0078 \pm 0.0050$	& $1.0537 \pm 0.0050$ &	 $0.0079002 \pm 0.0000020$ \\
g & 1 & $1.1067159 \pm 0.0000036$ & $0.5305 \pm 0.0050$ & $2.7944 \pm 0.0095$ &	$0.0078336 \pm 0.0000041$ \\
\hline
g & $-1$ & $1.1236835 \pm 0.0000022$ & $0.8880 \pm 0.0050$ & $2.9444 \pm 0.0057$ \\
g & 0 & $1.1315169 \pm 0.0000085$ & $0.2272 \pm 0.0050$	& $2.7564 \pm 0.0221$ &	$0.0078335 \pm 0.0000087$ \\
g & 1 & $1.1395122 \pm 0.0000159$ & $0.1210 \pm 0.0050$	& $-0.5877 \pm 0.0415$	& $0.0079953 \pm 0.0000180$ \\
\hline
g & $-1$ & $1.1585090 \pm 0.0000008$ & $2.5414 \pm 0.0050$ & $-1.8592 \pm 0.0020$ \\
g & 0 & $1.1664494 \pm 0.0000014$ & $1.4080 \pm 0.0050$	& $1.7753 \pm 0.0036$ &	 $0.0079404 \pm 0.0000016$ \\
g & 1 & $1.1743180 \pm 0.0000007$ & $2.5701 \pm 0.0050$	& $1.3201 \pm 0.0020$ &	 $0.0078686 \pm 0.0000016$ \\
\hline
g & $-1$ & $1.1956369 \pm 0.0000011$ & $1.7338 \pm 0.0050$ & $-2.3646 \pm 0.0029$ & \\
g & 0 & $1.2035735 \pm 0.0000009$ & $2.1082 \pm 0.0050$	& $-2.5097 \pm 0.0024$	& $0.0079366 \pm 0.0000014$ \\
g & 1 & $1.2114598 \pm 0.0000008$ & $2.3752 \pm 0.0050$	& $-2.3719 \pm 0.0021$ & $0.0078862 \pm 0.0000012$ \\
\hline
g & $-1$ & $1.2350651 \pm 0.0000067$ & $0.2839 \pm 0.0050$ & $2.5450 \pm 0.0177$ & \\
g & 0 & $1.2430074 \pm 0.0000019$ & $0.9890 \pm 0.0050$	& $0.9916 \pm 0.0051$ &	 $0.0079422 \pm 0.0000070$ \\
g & 1 & $1.2508969 \pm 0.0000050$ & $0.3869 \pm 0.0050$ & $1.2346 \pm 0.0130$ &	$0.0078895 \pm 0.0000053$ \\
\hline
g & $-1$ & $1.2775199 \pm 0.0000010$ & $1.9817 \pm 0.0050$ & $-2.0164 \pm 0.0025$ \\
g & 0 & $1.2854168 \pm 0.0000018$ & $1.0532 \pm 0.0050$ & $-0.6484 \pm 0.0048$ & $0.0078968 \pm 0.0000021$ \\
g & 1 & $1.2933148 \pm 0.0000008$ & $2.2686 \pm 0.0050$ & $-2.0449 \pm 0.0022$ & $0.0078980 \pm 0.0000020$ \\
\hline
g & $-1$ & $1.3230653 \pm 0.0000049$ & $0.3905 \pm 0.0050$ & $-2.7623 \pm 0.0129$ & \\
g & 0 & $1.3309425 \pm 0.0000074$ & $0.2582 \pm 0.0050$ & $0.4848 \pm 0.0194$ &	 $0.0078771 \pm 0.0000089$ \\
g & 1 & $1.3388459 \pm 0.0000055$ & $0.3496 \pm 0.0050$ & $0.1045 \pm 0.0143$ &	 $0.0079034 \pm 0.0000092$ \\
\hline
g & $-1$ & $1.3723898 \pm 0.0000055$ & $0.3472 \pm 0.0050$ & $-1.0984 \pm 0.0145$ & \\
g & 0 & $1.3803260 \pm  0.0000106$ & $0.1812 \pm 0.0050$ & $0.0662 \pm 0.0277$ & $0.0079362 \pm 0.0000119$ \\
g & 1 & $1.3881872 \pm 0.0000087$ & $0.2197 \pm 0.0050$ &  $2.7474 \pm 0.0228$ & $0.0078612 \pm 0.0000137$ \\
\hline
g & $-1$ & $1.4256857 \pm 0.0000053$ & $0.3595 \pm 0.0050$ & $-0.0756 \pm 0.0140$ & \\
g & 0 & $1.4336601 \pm 0.0000094$ & $0.2035 \pm 0.0050$ & $2.1476 \pm 0.0246$ &	 $0.0079744 \pm 0.0000108$ \\
g & 1 & $1.4415400 \pm 0.0000056$ & $0.3416 \pm 0.0050$ & $1.7827 \pm 0.0147$ & $0.0078799 \pm 0.0000109$  \\
\hline
g & $-1$ & $1.4834010 \pm 0.0000025$ & $0.7541 \pm 0.0050$ & $-1.7034 \pm 0.0067$ & \\
g & 0 & $1.4913387 \pm 0.0000041$ & $0.4706 \pm 0.0050$ & $-1.2904 \pm 0.0107$ & $0.0079377 \pm 0.0000048$ \\
g & 1 & $1.4992400 \pm 0.0000083$ & $0.2315 \pm 0.0050$ & $2.3651 \pm 0.0217$ &	 $0.0079012 \pm 0.0000092$ \\
\hline
g & $-1$ & $1.5463993 \pm 0.0000133$ & $0.1435 \pm 0.0050$ & $-0.8828 \pm 0.0350$ & \\
g & 0 & $1.5543411 \pm 0.0000081$ & $0.2358 \pm 0.0050$ & $-2.5694 \pm 0.0213$ &	 $0.0079417 \pm 0.0000156$ \\
g & 1 & $1.5623302 \pm 0.0000310$ & $0.0618 \pm 0.0050$	& $-2.6040 \pm 0.0813$ & $0.0079891 \pm 0.0000320$ \\
\hline
g & $-1$ & $1.6887614 \pm 0.0000184$ & $0.1032 \pm 0.0050$ & $2.6632 \pm 0.0486$ & \\
g & 0 & $1.6967366 \pm 0.0000166$ & $0.1152 \pm 0.0050$ & $-1.9493 \pm 0.0435$ &	 $0.0079752 \pm 0.0000248$ \\
g & 1 & $1.7046912 \pm 0.0000124$ & $0.1545 \pm 0.0050$ & $-1.7011 \pm 0.0325$ &	 $0.0079546 \pm 0.0000207$ \\
\hline

g & $-1$ & $1.7692910 \pm 0.0000481$ & $0.0397 \pm 0.0050$ & $2.9018 \pm 0.1264$ & \\
g & 0 & $1.7771923 \pm 0.0000658$ & $0.0291 \pm 0.0050$ & $2.8920 \pm 0.1725$ &	 $0.0079013 \pm 0.0000815$ \\
g & 1 & $1.7850883 \pm 0.0000231$	& $0.0825 \pm 0.0050$ & $-1.1420 \pm 0.0608$ &	 $0.0078961 \pm 0.0000697$ \\
\hline
\hline
\end{tabular}
\label{table:9244992_g}
\end{table*}

\begin{table*}
\centering
\caption[]{
Additional  g-mode frequencies of  KIC~9244992. }
\begin{tabular}{crccrc}
\hline
\hline
& $m$ & \multicolumn{1}{c}{frequency} & \multicolumn{1}{c}{amplitude} &
\multicolumn{1}{c}{phase} & \multicolumn{1}{c}{$\Delta \nu_{\rm rot}$} \\
& & \multicolumn{1}{c}{d$^{-1}$} & \multicolumn{1}{c}{mmag} &
\multicolumn{1}{c}{radians} & \multicolumn{1}{c}{d$^{-1}$} \\
\hline
 g & ? &  $0.4251423 \pm 0.0000208$ & $0.0913 \pm 0.0050$ & $-2.7407 \pm 0.0548$ & \\	
 g & ? &  $	0.4561535	\pm 0.0000111$ & $0.1712	\pm 0.0050$  & $2.2152 \pm 0.0292$ &	$0.0310112 \pm 0.0000236$ \\	
 g & ? & $0.4874012 \pm 0.0000184$ & $0.1032	\pm 0.0050$ & $1.1203 \pm 0.0485$ &	$0.0312477 \pm 0.0000215$ \\
\hline
g & ? & $1.1472959 \pm 0.0000125$ & $0.1527 \pm 0.0050$	& $1.9426 \pm 0.0329$ &	$0.0077837 \pm 0.0000202$ \\
\hline
g & ? & $1.5161600 \pm 0.0000087$ & $0.2194 \pm 0.0050$	& $-0.0377 \pm 0.0228$ & \\
\hline
 g & -1 &	$4.7813912 \pm 0.0000689$ & $0.0277 \pm 0.0050$ &  $1.2119 \pm 0.1811$  & \\ 
 g & 0 & $4.7892914	\pm 0.0000265$ & $0.0719	\pm 0.0050$  & $2.0760 \pm 0.0697$ & $0.0079003 \pm 0.0000738$	\\
 g & 1 & $4.7972105	\pm 0.0000404$ & $0.0471	\pm 0.0050$  &	$-2.2399 \pm 0.1063$ &	 $0.0079190 \pm 0.0000483$ \\
\hline
\hline
\end{tabular}
\label{table:9244992_g_add}
\end{table*}

\begin{figure}
\centering
\includegraphics[width=0.95\linewidth,angle=0]{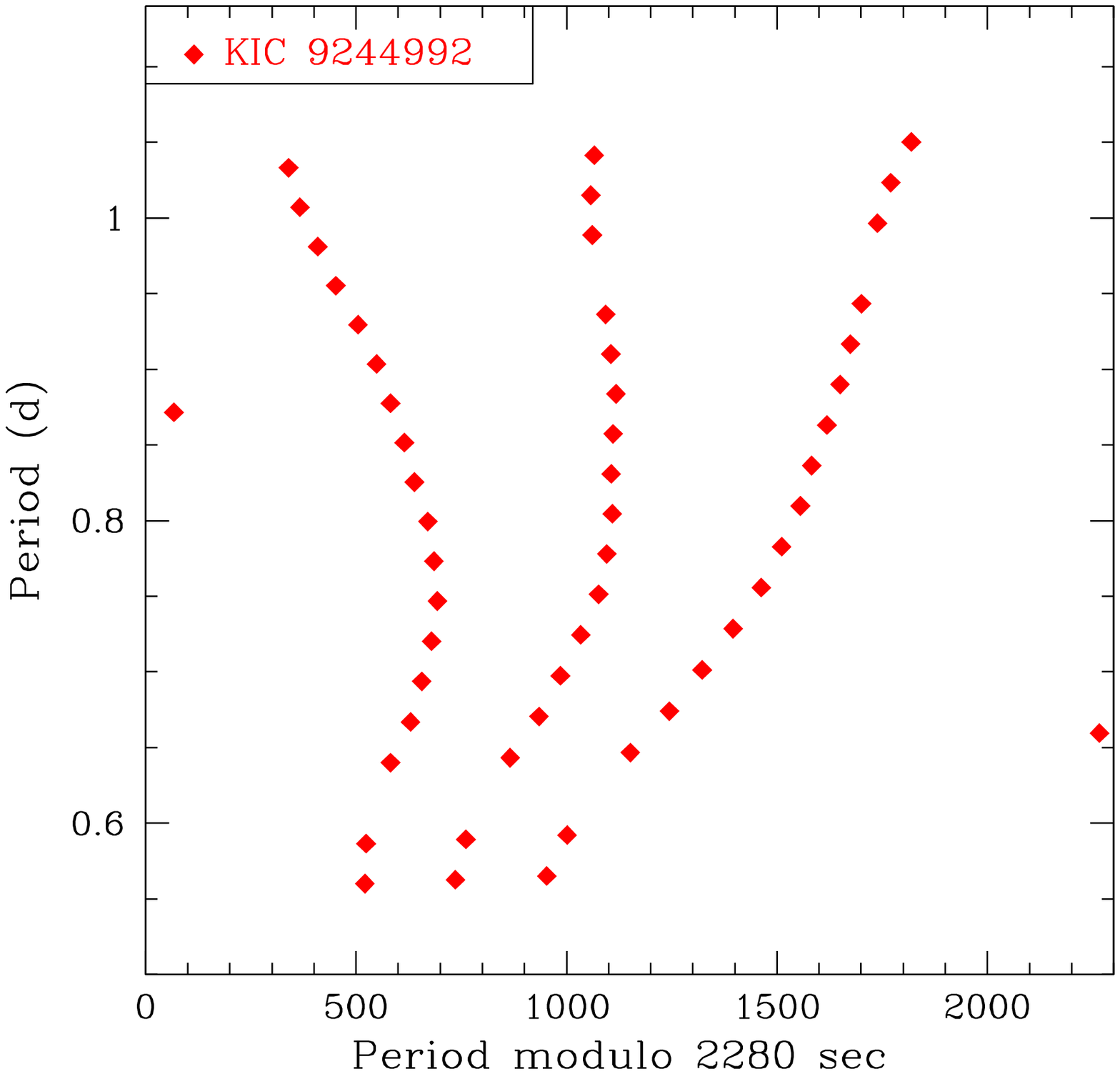} 
\caption{The \'echelle diagram for all the g~modes (except the longest period triplet) detected in KIC~9244992. 
Most of the frequencies in the g-mode range are dipole modes that form triplets; the periods of each component of the multiplets, i.e., $m = -1, 0$ or $+1$, are  nearly equally spaced.
}
\label{fig:k9244992echelle}
\end{figure}

The periods corresponding to all frequencies (except for the triplets at
$2.09$~d$^{-1}$ and $0.456$~d$^{-1}$) 
obtained in the g-mode range (Tables~\ref{table:9244992_g} and \ref{table:9244992_g_add}) are plotted against the period modulus of 2280~s ($0.0264$~d)   in Fig.~\ref{fig:k9244992echelle} \citep[see also][]{bedding2014}. This \'echelle diagram confirms that the g-mode triplets (i.e., $l=1$ modes) are spaced nearly equally in period, with some modulation. Except for a few modes, we have detected most of the consecutive radial orders of g~modes between $0.95$~d$^{-1}$ and  $1.78$~d$^{-1}$.  Two outliers at periods of  $0.66$~d (frequency $1.516$~d$^{-1}$)  and $0.87$~d (frequency $1.147$~d$^{-1}$) in Fig.~\ref{fig:k9244992echelle} might belong to quadrupole ($l=2$) modes, but their mode identification is not certain.

\subsection{The p~modes}
\label{sec:pmodes}

\begin{figure}
\centering
\includegraphics[width=0.9\linewidth,angle=0]{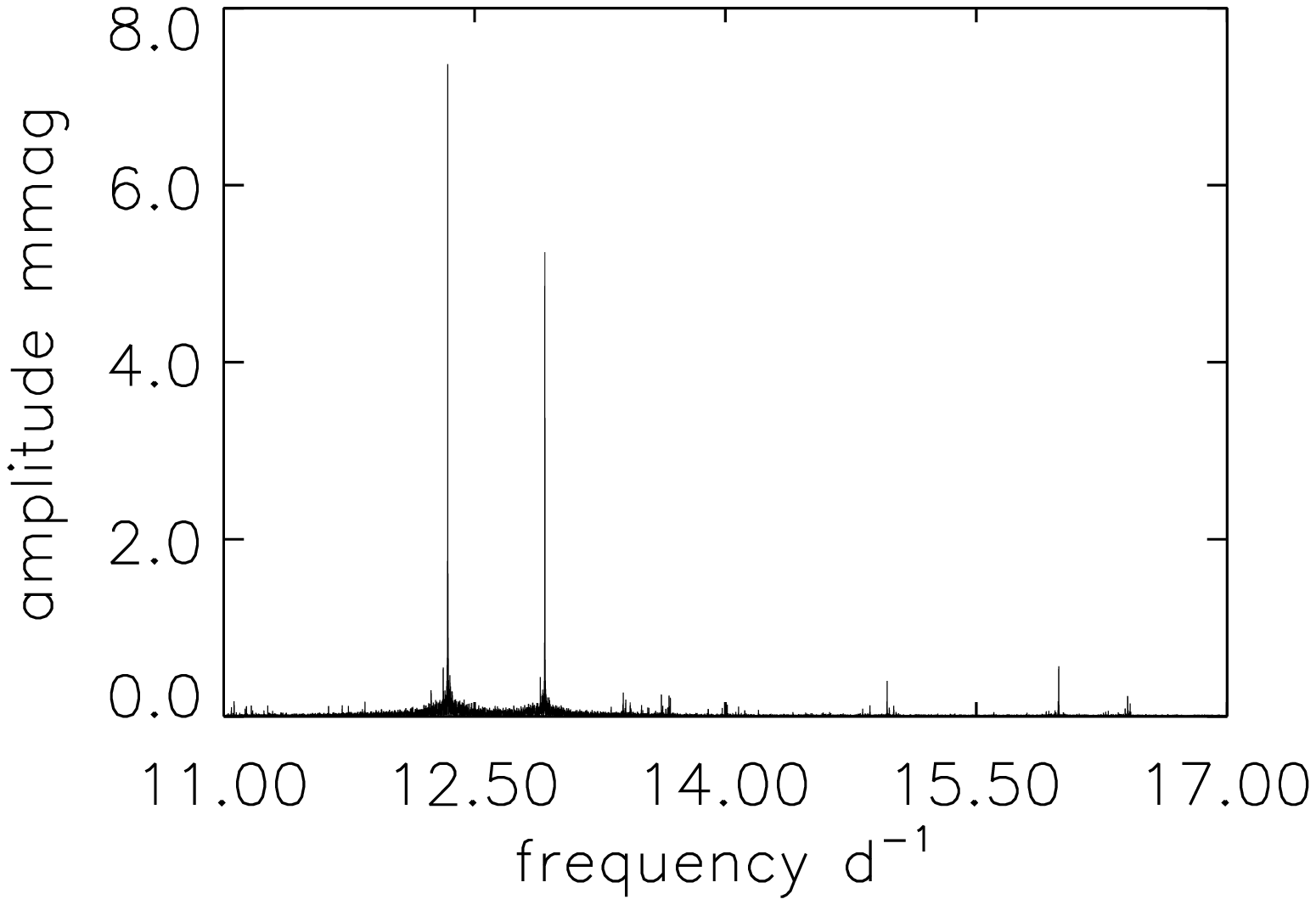}\\ 
\includegraphics[width=0.9\linewidth,angle=0]{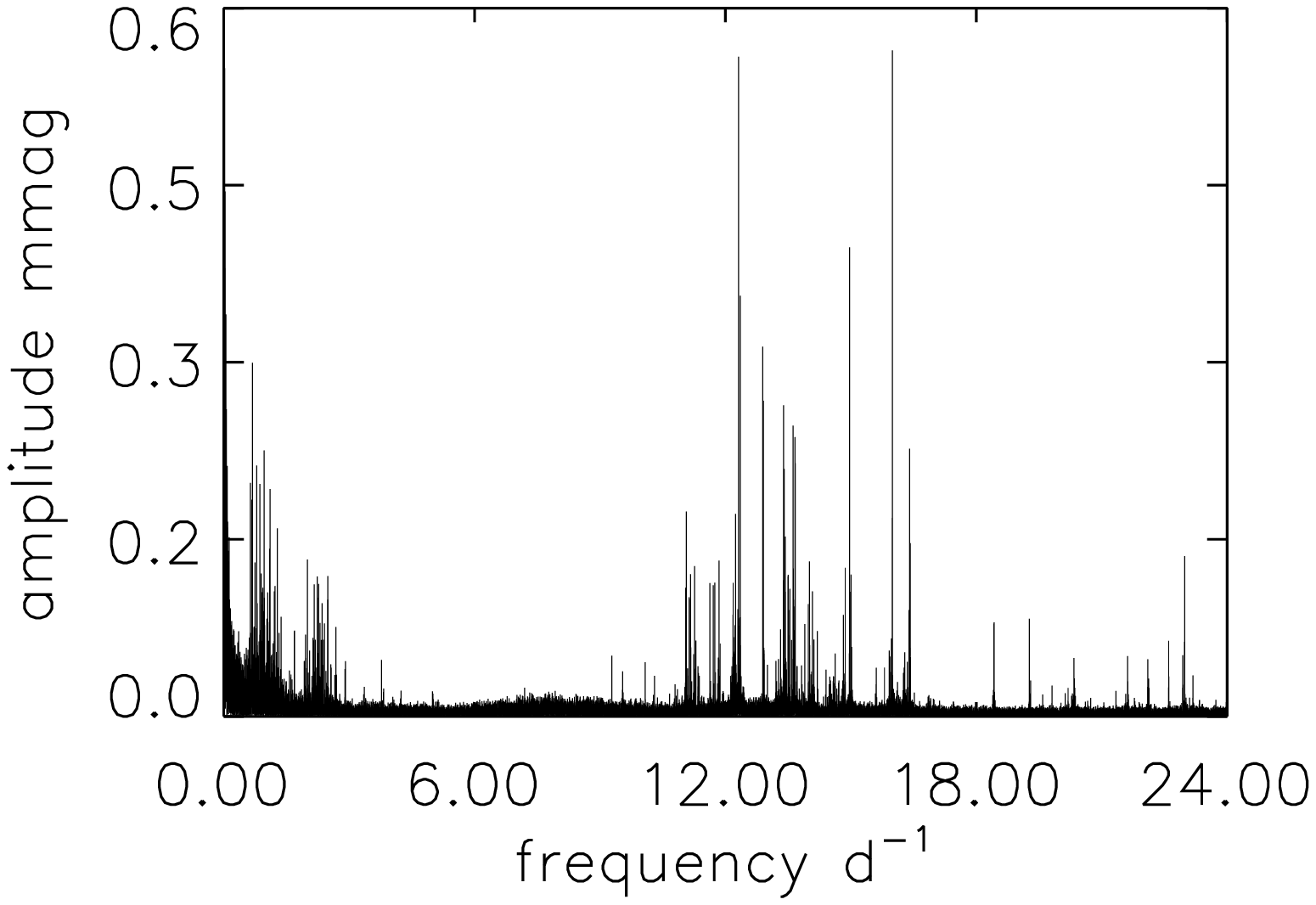}   
\caption{
Top panel: The amplitude spectrum for the p-mode range of $11 - 17$~d$^{-1}$ where most of the p-modes peaks are. The two high amplitude peaks, named $\nu_1$ and $\nu_2$ in the order of amplitude,  are obvious. Bottom: The full frequency range after pre-whitening by $\nu_1$ and $\nu_2$ where a plethora of peaks can be seen.
}
\label{fig:9244992_ftp1}
\end{figure}

The top panel of Fig.~\ref{fig:9244992_ftp1} shows the amplitude spectrum in the frequency range of the highest amplitude p-mode peaks. There are two dominant peaks, at $12.339$~d$^{-1}$ ( $=\nu_1$) and at $12.920$~d$^{-1}$ ($=\nu_2$). These two peaks are prewhitened in the amplitude spectrum in the bottom of Fig.~\ref{fig:9244992_ftp1}, where many peaks in the p-mode range can now be seen. Frequencies, amplitudes and phases of the most significant peaks are listed in Tables~\ref{table:9244992_p} and \ref{table:9244992_p_add}.

\begin{figure}
\centering
\includegraphics[width=0.9\linewidth,angle=0]{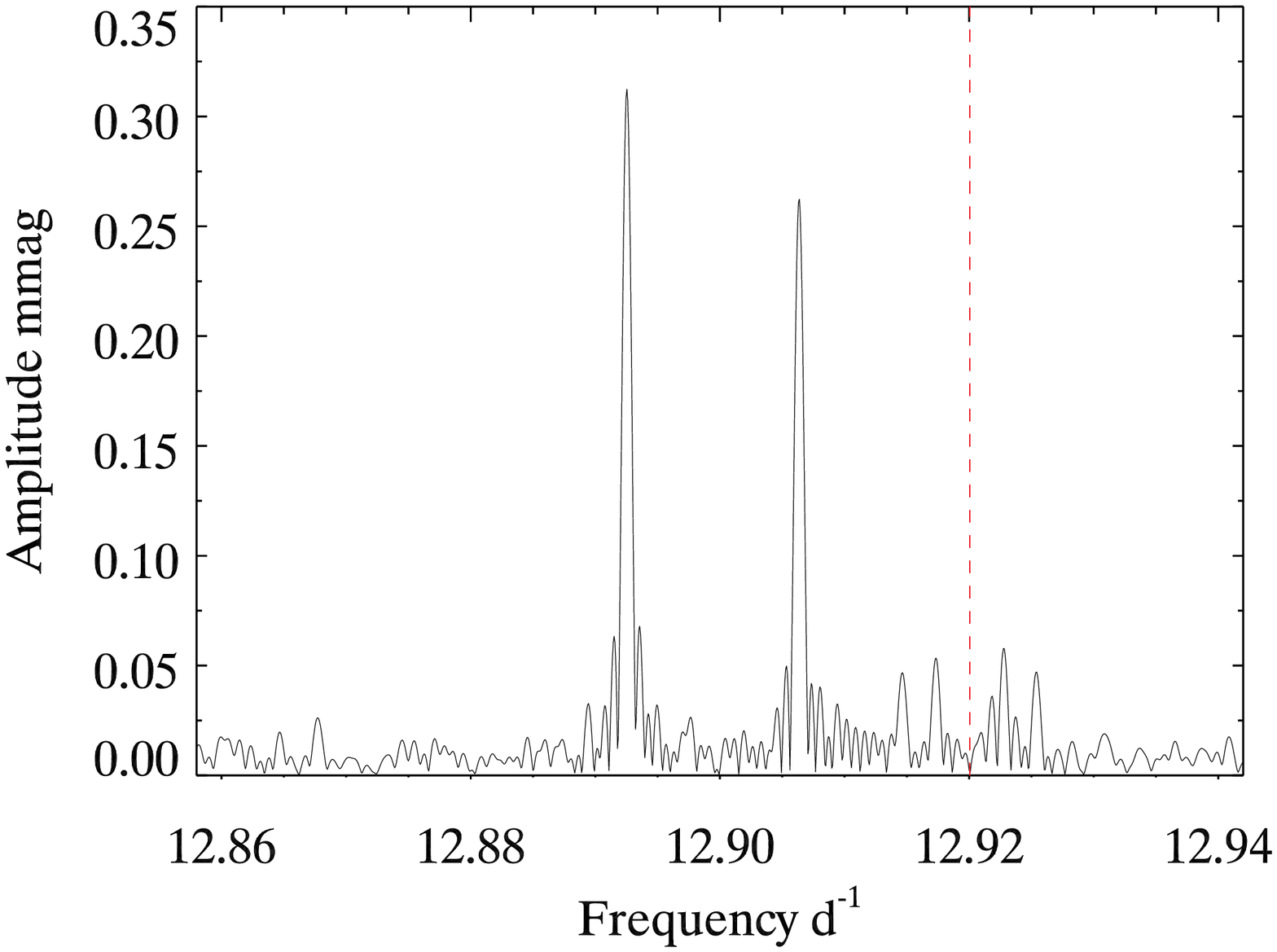}\\ 
\includegraphics[width=0.9\linewidth,angle=0]{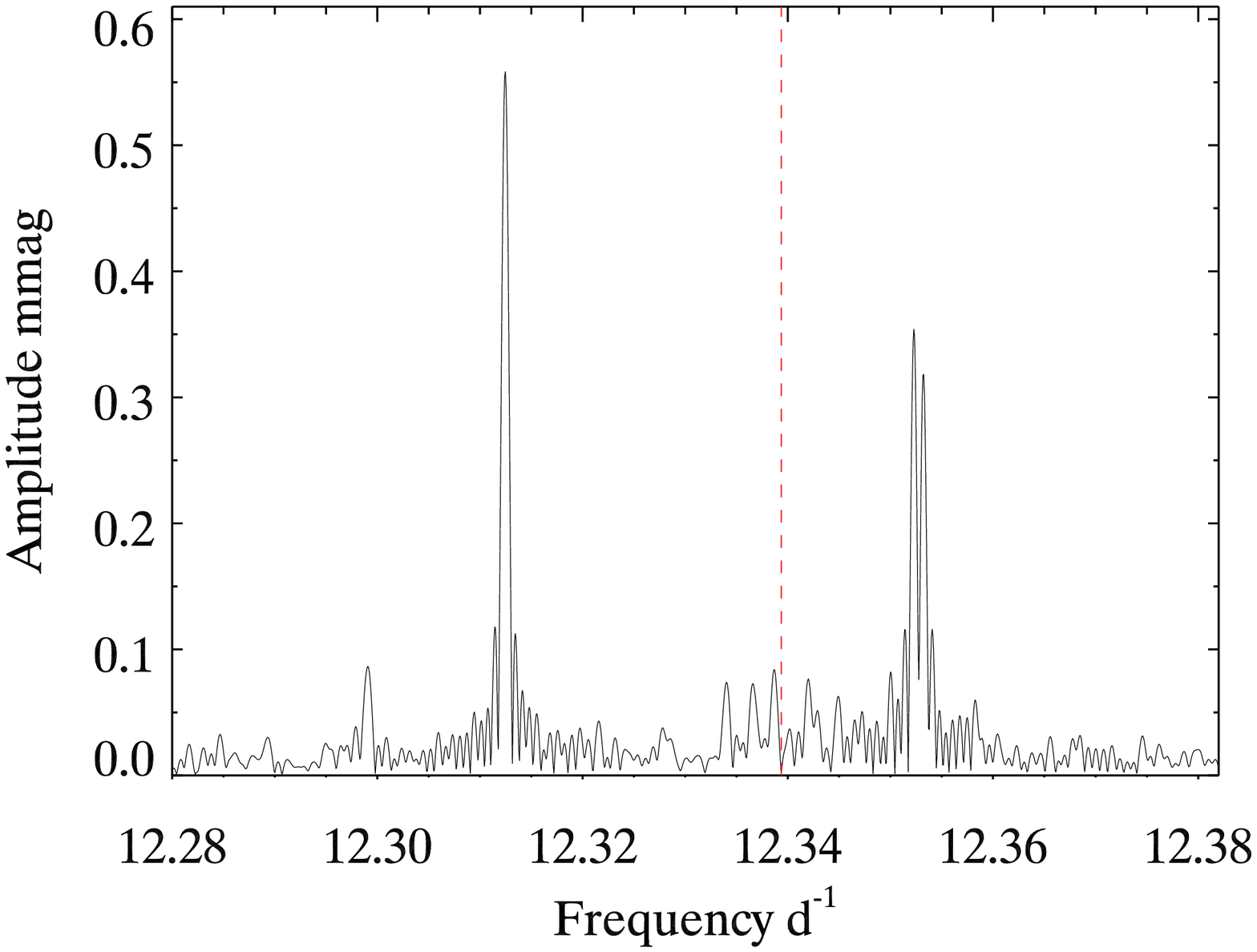}   
\caption{
{\bf Top:} The range around $\nu_2 = 12.920$~d$^{-1}$ after prewhitening $\nu_2$ (dashed line). The two highest peaks are the $m=-1$ (left) and $m=0$ (right) components of the triplet formed with $\nu_2$. Where $\nu_2$ has been prewhitened
(dashed line) there are four low amplitude peaks separated from $\nu_2$ by the Kepler orbital frequency, hence these are artefacts. 
{\bf Bottom:} The range around $\nu_1=12.339$~d$^{-1}$ after prewhitening $\nu_1$ (dashed lines). There are no obvious rotationally split components, which supports the identification of $\nu_1$ as a radial mode. The  relation of the two peaks at
$12.312$~d$^{-1}$ and $12.352$~d$^{-1}$ is not clear.
}
\label{fig:9244992_ftp2}
\end{figure}

\begin{figure}
\centering
\includegraphics[width=0.9\linewidth,angle=0]{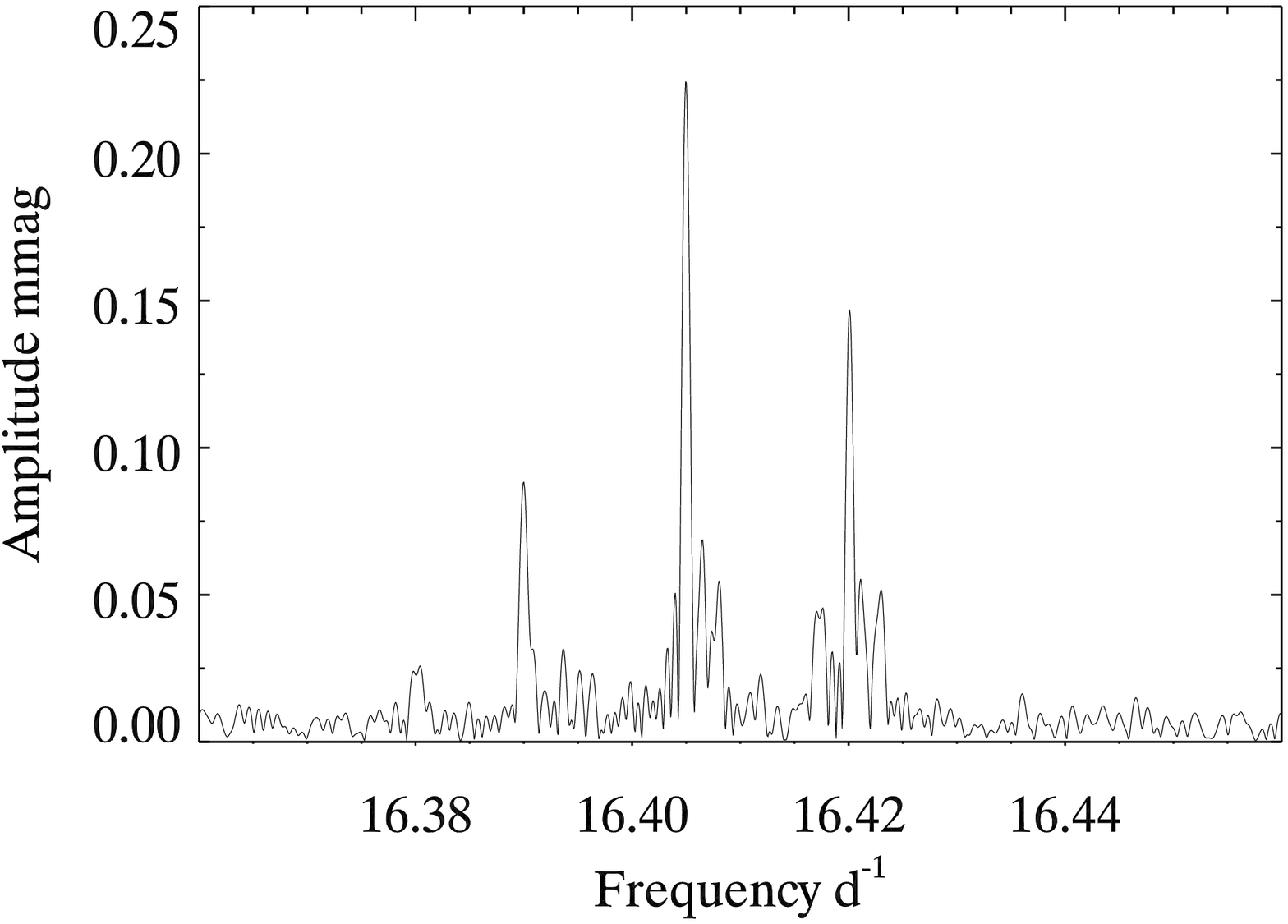}\\ 
\includegraphics[width=0.9\linewidth,angle=0]{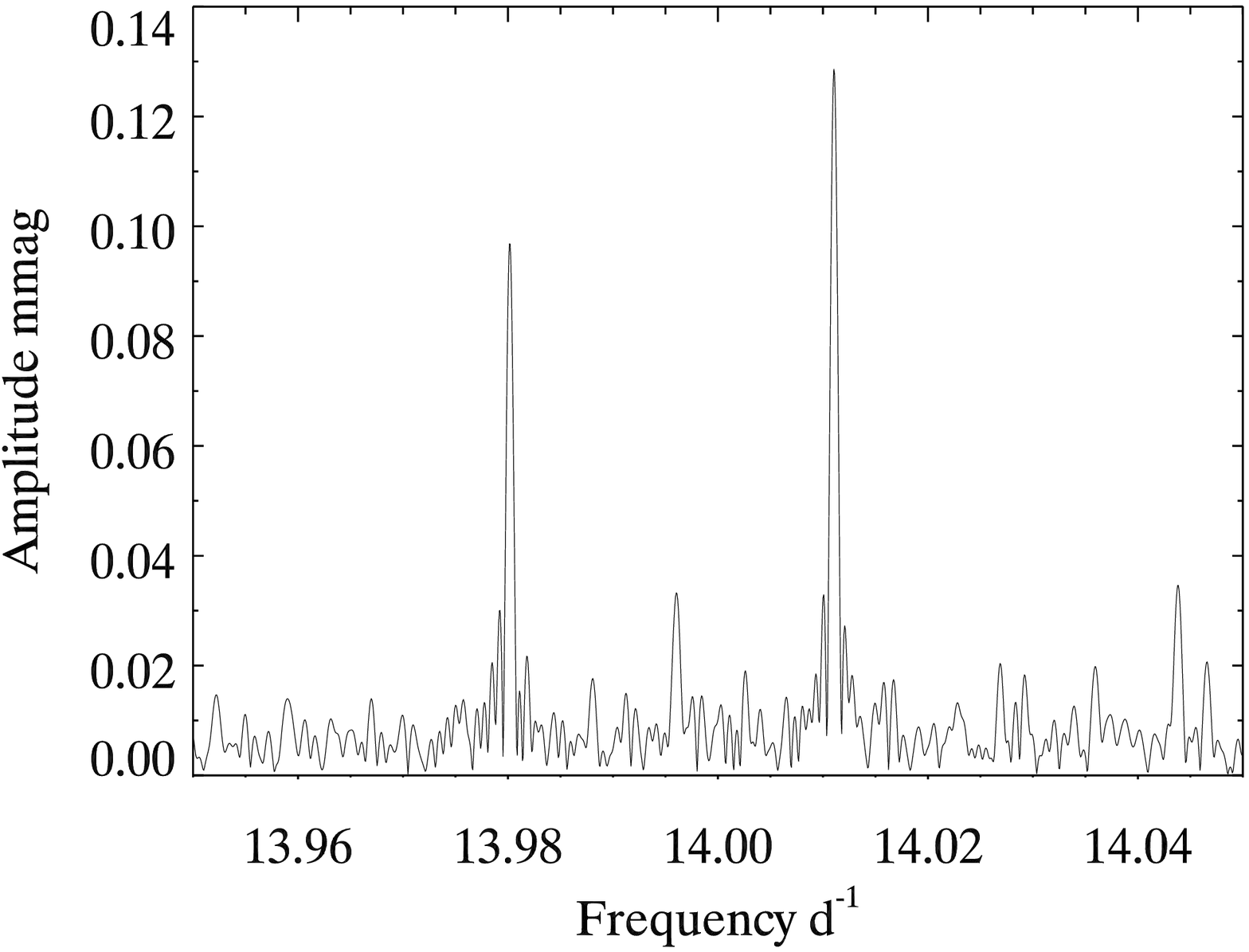}   
\caption{Amplitude spectra around two triplets at $\nu_6=16.405$~d$^{-1}$ (top panel) and $\nu_7=13.996$~d$^{-1}$ (bottom panel).
}
\label{fig:9244992_ftp3}
\end{figure}

The top panel of Fig.~\ref{fig:9244992_ftp2} shows the region around $\nu_2$ after it has been prewhitened (dashed line). We identify the two highest peaks as the $m=-1$ (left) and $m=0$ (right) components of the triplet formed with $\nu_2$ ($m=1$). The bottom panel shows the region around $\nu_1$, which has been prewhitened (dashed line). No obvious rotationally split components are found, suggesting $\nu_1$ to be a radial mode. There are peaks at $12.312$~d$^{-1}$ and $12.352$~d$^{-1}$, which might be members ($m=-1$ and $+2$) of a quintuplet with a  central frequency of $\approx 12.325$~d$^{-1}$. Both frequencies are labelled as $\nu_5$ (Table~\ref{table:9244992_p_add}). Fig.~\ref{fig:9244992_ftp3} shows another two triplets labelled $\nu_6$ and $\nu_7$. Looking at all these dipole triplets, the height of the central peak ($m=0$) relative to the other peaks varies.
Hence, the amplitude ratio cannot be used to infer the inclination of the pulsation axis.  

\begin{figure}
\centering
\includegraphics[width=0.9\linewidth,angle=0]{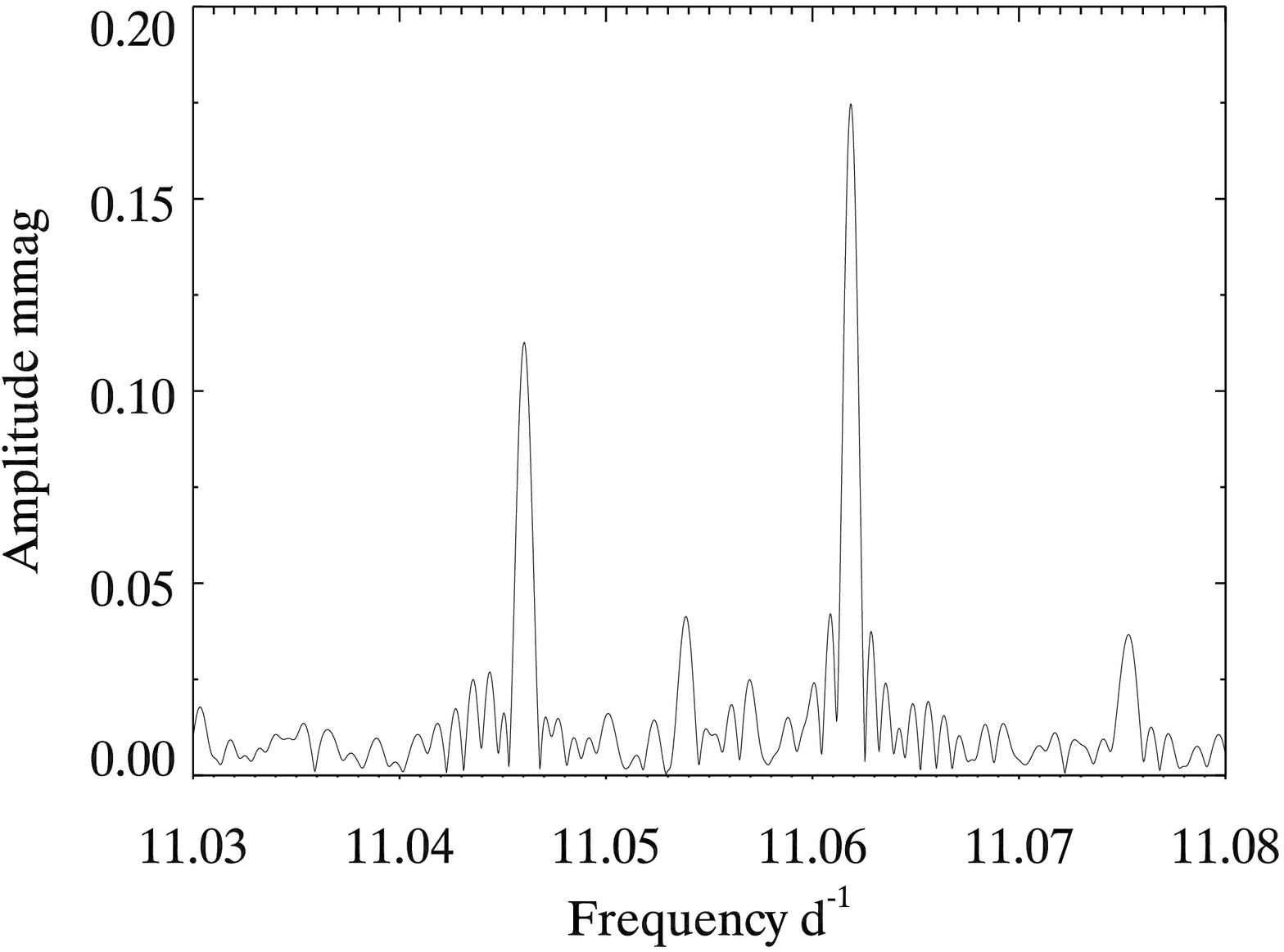}\\ 
\includegraphics[width=0.9\linewidth,angle=0]{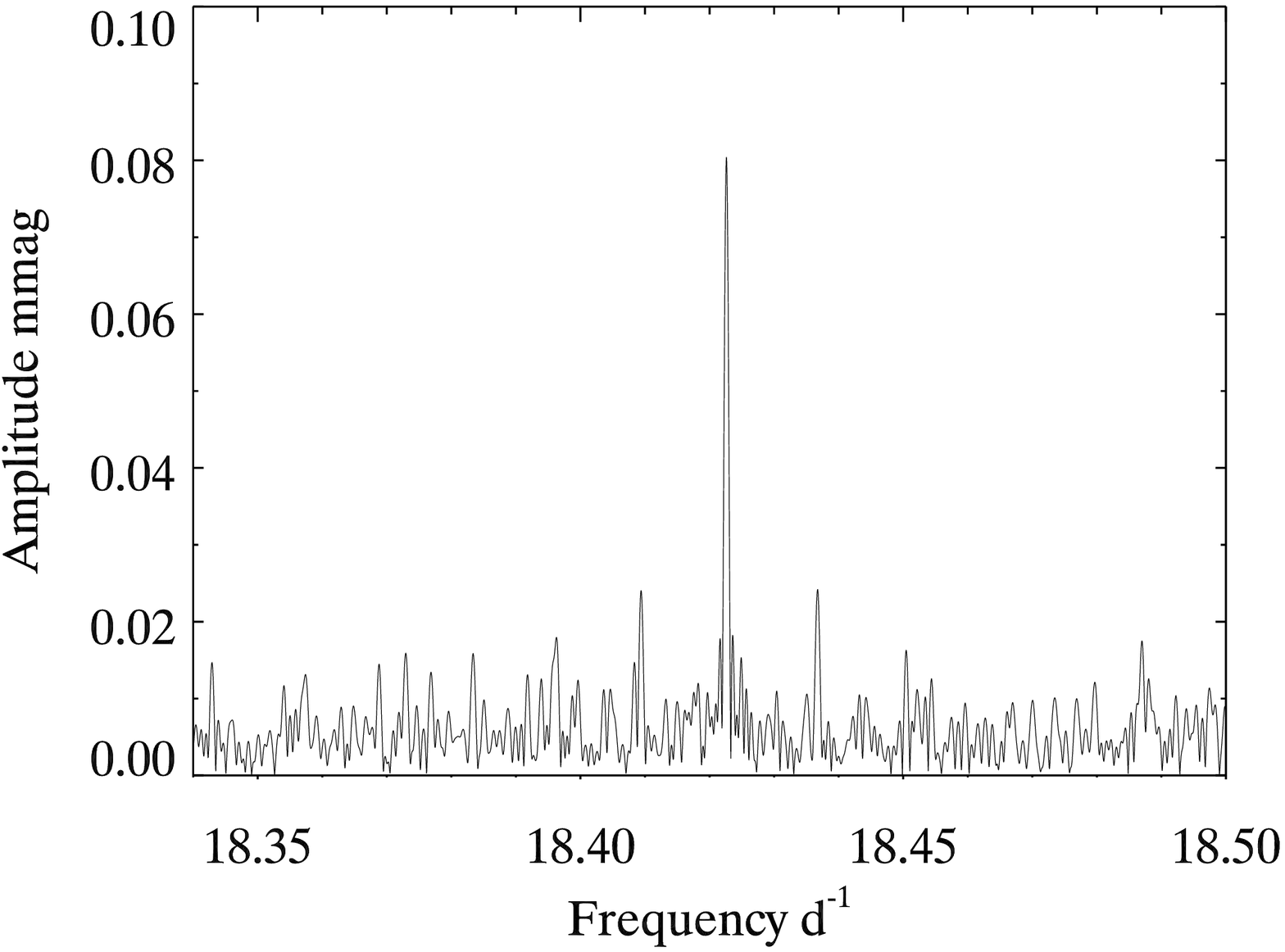}   
\caption{
Top panel: An amplitude spectrum showing a triplet in the p-mode range split by the same amount as the g~modes. This triplet has frequencies that are the difference between $\nu_1 = 12.339368$~d$^{-1}$ and the g-mode triplet given in  Table~\ref{table:9244992_g} centred on $\nu = 1.2854168$~d$^{-1}$. Therefore, these triplet frequencies are cross-terms showing that the p~modes and g~modes are coupled, just as in KIC~11145123 \citep{kurtz14}. Bottom panel: An isolated p-mode quintuplet (or septuplet). 
}
\label{fig:9244992_ftp4}
\end{figure}

The top panel of Fig.~\ref{fig:9244992_ftp4} shows another triplet in the p-mode range. Since the splitting among the components is the same as those of g-mode triplets, the three frequencies of this triplet must be cross-terms caused by coupling between the p~modes and g~modes; similar phenomena are also found in KIC~11145123 \citep{kurtz14}. The presence of the cross-terms confirms that the g~modes and the p~modes originate in the same star. The bottom panel of  Fig.~\ref{fig:9244992_ftp4} shows an isolated quintuplet (or possibly septuplet), showing that higher degree, $l$, modes are excited. 

\begin{table*}
\centering
\caption[]{
This Table shows the singlet and triplet frequencies in the p-mode range of KIC~9244992.  }
\begin{tabular}{ccrccrc}
\hline
&& &\multicolumn{1}{c}{frequency} & \multicolumn{1}{c}{amplitude} &
\multicolumn{1}{c}{phase} & \multicolumn{1}{c}{$\Delta \nu_{\rm rot}$}
 \\
&& $m$ &\multicolumn{1}{c}{d$^{-1}$} & \multicolumn{1}{c}{mmag} &
\multicolumn{1}{c}{radians} & \multicolumn{1}{c}{d$^{-1}$} \\
\hline
\hline
$\nu_1$ &s & 0 & $12.3393680 \pm 0.0000003$ & $7.3701 \pm 0.0050$ & $2.0083 \pm 0.0007$ \\
\hline
&  & $-1$ & $12.8925272 \pm 0.0000060$	& $0.3156 \pm 0.0050$ & $-0.8012 \pm 0.0158$ \\
$\nu_2$ & t  & 0 & $12.9063656 \pm 0.0000070$ & $0.2704 \pm 0.0050$ & $1.8549 \pm 0.0185$ & $0.0138383 \pm 0.0000093$ \\
&  & 1 & $12.9200576 \pm 0.0000004$ & $5.2472 \pm 0.0050$ & $1.6134 \pm 0.0010$ & $0.0136920 \pm 0.0000070$ \\
\hline
&  & $-1$ & $13.9801610 \pm 0.0000206$ & $0.0924 \pm 0.0050$ & $2.4194 \pm 0.0541$ \\
$\nu_7$ & t & 0 & $13.9959239 \pm 0.0000695$ & $0.0274 \pm 0.0050$ & $0.6543 \pm 0.1827$ & $0.0157629 \pm 0.0000725$ \\
&  & 1 & $14.0110463 \pm 0.0000145$ & $0.1310 \pm 0.0050$ & $-3.0844 \pm 0.0381$	 & $0.0151225 \pm 0.0000710$ \\
\hline
&  & $-1$ & $14.0437670 \pm 0.0000561$ & $0.0340 \pm 0.0050$ & $0.9505 \pm 0.1475$ \\
$\nu_8$ & t & 0 & $14.0617857 \pm 0.0000530$ & $0.0360 \pm 0.0050$ & $40.4102 \pm 0.1394$ & $0.0180187 \pm 0.0000771$ \\
&  & 1 & $14.0785403 \pm 0.0000179$ & $0.1058 \pm 0.0050$ & $2.3197 \pm 0.0472$ & $0.0167546 \pm 0.0000559$ \\
\hline
&  & $-1$ & $14.7916430 \pm 0.0000826$	& $0.0230 \pm 0.0050$ & $-1.1759 \pm 0.2173$ \\
$\nu_9$ & t & 0 & $14.8060543 \pm 0.0000651$ & $0.0293 \pm 0.0050$ & $-1.0657 \pm 0.1712$ & $0.0144114 \pm 0.0001052$ \\
&  & 1 & $14.8204663 \pm 0.0000223$ & $0.0850 \pm 0.0050$ & $2.0742 \pm 0.0587$ & $0.0144120 \pm 0.0000688$ \\
\hline
&  & $-1$ & $15.5900915 \pm 0.0001015$	& $0.0188 \pm 0.0050$ & $1.4045	\pm 0.2673$ \\
$\nu_{11}$ & t & 0 & $15.6048261 \pm 0.0000470$ & $0.0403 \pm 0.0050$ & $-2.6684\pm 0.1238$ & $0.0147346 \pm 0.0001119$ \\
&  & 1 & $15.6193748 \pm 0.0001209$ & $0.0157 \pm 0.0050$ & $-1.5799 \pm 0.3185$ & $0.0145487 \pm 0.0001298$ \\
\hline
&  & $-1$ & $16.3899664 \pm 0.0000219$ & $0.0868 \pm 0.0050$ & $-2.4682 \pm 0.0575$ \\
$\nu_6$ & t & 0 & $16.4049834 \pm 0.0000084$ & $0.2266 \pm 0.0050$ & $0.9131 \pm 0.0220$ & $0.0150169 \pm 0.0000234$ \\
&  & 1 & $16.4200629 \pm 0.0000130$ & $0.1458 \pm 0.0050$ & $0.4931 \pm 0.0342$ & $0.0150795 \pm 0.0000155$ \\
\hline
&  & -3 &	$18.3831600 \pm 0.0001959$ & $0.0096 \pm 0.0050$ & $-2.2904 \pm 0.5157$ & \\
&  & -2 & $18.3962411 \pm 0.0001181$ & $0.0161 \pm 0.0050$ & $-0.8998 \pm 0.3107$	 & $0.0130811 \pm 0.0002287$ \\
&  & -1 & $18.4093327 \pm 0.0000842$ & $0.0225 \pm 0.0050$ & $2.3136 \pm 0.2214$ & $0.0130915 \pm 0.0001450$ \\
$\nu_{10}$ &sep?  & 0 & $18.4225992 \pm 0.0000240$ & $0.0793 \pm 0.0050$ & $0.3251 \pm 0.0632$ & $0.0132665 \pm 0.0000876$ \\
&  & 1 & $18.4366948 \pm 0.0000914$ & $0.0207 \pm 0.0050$ & $1.5643 \pm 0.2404$ & $0.0140956 \pm 0.0000945$ \\
&  & 2 & $18.4521563 \pm 0.0001611$ & $0.0117 \pm 0.0050$ & $2.5397 \pm 0.4235$ & $0.0154615 \pm 0.0001852$ \\	
\hline
\hline
\end{tabular}
\label{table:9244992_p}
\end{table*}

\begin{table*}
\centering
\caption[]{
Additional frequencies in the p-mode range of KIC~9244992 whose multiplicities are not clear.  }
\begin{tabular}{cccr}
\hline
& \multicolumn{1}{c}{frequency} & \multicolumn{1}{c}{amplitude} &
\multicolumn{1}{c}{phase}  \\
& \multicolumn{1}{c}{d$^{-1}$} & \multicolumn{1}{c}{mmag} &
\multicolumn{1}{c}{radians} \\
\hline
\hline
$\nu_5$ &  $12.3124621 \pm 0.0000034$ & $0.5625 \pm 0.0050$ & $-2.0569 \pm 0.0089$\\
$\nu_5$ &  $12.3522912 \pm 0.0000053$ & $0.3604 \pm 0.0050$ & $-1.4217 \pm 0.0139$\\
\hline
$\nu_4$ &  $14.9664768 \pm 0.0000048$ & $0.3993 \pm 0.0050$ & $-2.9340 \pm 0.0126$\\
$\nu_3$ &  $15.9929674 \pm 0.0000043$ & $0.5646 \pm 0.0050$ & $-0.4456  \pm 0.0089$\\
\hline
\hline
\end{tabular}
\label{table:9244992_p_add}
\end{table*}

Frequencies obtained in the p-mode range are listed in Tables~\ref{table:9244992_p} and \ref{table:9244992_p_add}. The highest amplitude singlet and well-resolved triplet frequencies are given in Table~\ref{table:9244992_p}, while additional frequencies are given in Table~\ref{table:9244992_p_add}.
Rotational splittings, $\Delta\nu_{\rm rot}$, of p-mode triplets are about twice the mean splittings of g~modes. This indicates that the internal rotation of KIC~9244992 is nearly uniform. We discuss this in more detail in section~\ref{sec:internal_rotation}.

\section{Model}
\label{sec:model}

In this section we discuss the mass and the evolutionary stage of KIC~9244992, by comparing observed pulsation frequencies with those obtained from evolutionary models with various parameters. The models were calculated in the same way as by \citet{kurtz14}, using the Modules for Experiments in Stellar Evolution (MESA) code \citep{paxton2013}.  For KIC~9244992 two important facts can be used to constrain the model: the period spacing of the high-order g~modes, and the frequency of a radial pulsation mode, $\nu_1$.

\subsection{The period spacing of the g~modes and the radial mode $\nu_1$}

Table~\ref{table:periods} lists the frequency, radial order $n$, and the period of the central mode ($m=0$) of each detected g-mode triplet, where the radial order $n$ is adopted from our best model (see below). The last column gives the period spacing from the adjacent $(n-1)$ mode, if available. The period spacings in slowly rotating stars are nearly constant, which is a well-known character of high-order g~modes \citep{unno1989,aerts2010}.  Note that in more rapidly rotating $\gamma$ Dor stars, this is not the case \citep{bedding2014,bouabid2013,vanreeth2014}. 

\begin{table}
\centering
\caption[]{The central g-mode frequencies for each triplet.  The second column then gives the corresponding period, and the final column gives the period spacing to the previous  mode. 
}
\begin{tabular}{rccc}
\hline
\multicolumn{1}{c}{$n$}
 &
\multicolumn{1}{c}{frequency} & \multicolumn{1}{c}{period (P)} &
\multicolumn{1}{c}{$P_{n-1}-P_{n}$} \\
&
\multicolumn{1}{c}{d$^{-1}$} & \multicolumn{1}{c}{d} &
\multicolumn{1}{c}{d} \\
\hline
\hline
$-38$&$0.960153$ & $1.041501 \pm 0.000007$  \\
$-37$&$0.985201$ & $1.015022 \pm 0.000002$ & $0.026479 \pm 0.000008$ \\
$-36$&$1.011456$ & $0.988673 \pm 0.000018$ & $0.026348 \pm 0.000018$\\
\\
$-34$&$1.068073$ & $0.936265 \pm 0.000002$  \\
$-33$&$1.098882$ & $0.910016 \pm 0.000002$ & $0.026250 \pm 0.000002$ \\
$-32$&$1.131517$ & $0.883769 \pm 0.000007$ & $0.026246 \pm 0.000007$ \\
$-31$&$1.166449$ & $0.857302 \pm 0.000001$ & $0.026467 \pm 0.000007$ \\
$-30$&$1.203573$ & $0.830859 \pm 0.000001$ & $0.026443 \pm 0.000001$ \\
$-29$&$1.243007$ & $0.804500 \pm 0.000001$ & $0.026359 \pm 0.000001$ \\
$-28$&$1.285417$ & $0.777958 \pm 0.000001$ & $0.026543 \pm 0.000002$ \\
$-27$&$1.330942$ & $0.751347 \pm 0.000004$ & $0.026611 \pm 0.000004$ \\
$-26$&$1.380326$ & $0.724467 \pm 0.000005$ & $0.026881 \pm 0.000007$ \\
$-25$&$1.433660$ & $0.697515 \pm 0.000004$ & $0.026951 \pm 0.000007$ \\
$-24$&$1.491339$ & $0.670538 \pm 0.000002$ & $0.026977 \pm 0.000005$ \\
$-23$&$1.554341$ & $0.643359 \pm 0.000003$ & $0.023875 \pm 0.000006$\\
\\
$-21$&$1.696737$ & $0.589367 \pm 0.000006$  \\
$-20$&$1.777192$ & $0.562685 \pm 0.000021$ & $0.026681 \pm 0.000022$\\ 
\\
$-7$&$4.789291$ & $0.208799 \pm 0.000001$ \\
\hline
\hline
\end{tabular}
\label{table:periods}
\end{table}

\begin{figure}
\centering
\includegraphics[width=0.98\linewidth,angle=0]{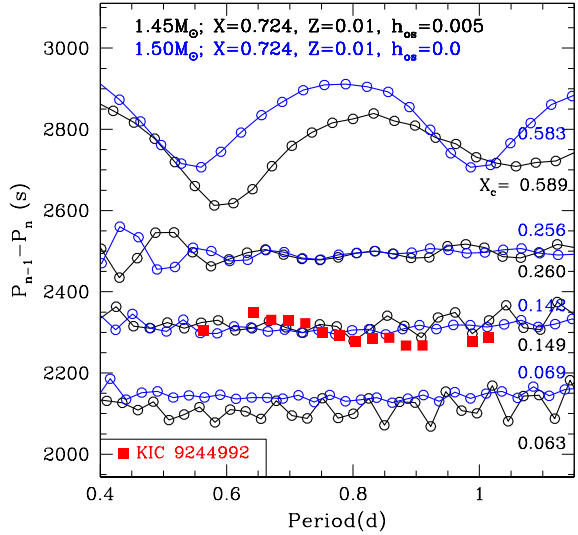} 
\caption{Period spacings of g~modes ($P_{n-1}-P_n$) as a function of g-mode period in selected evolutionary states marked by core hydrogen abundances. Observed spacings are compared with 1.45~M$_\odot$ and 1.50~M$_\odot$ models (a small overshooting is included in the former model). The observed spacings are consistent with evolved models with the central hydrogen abundance reduced to as little as $\sim$0.1.
}
\label{fig:Dpg_comp}
\end{figure}

Period spacings, $P_{n-1}-P_n$, of high-order g~modes are useful to determine the evolutionary stage of the star. Fig.~\ref{fig:Dpg_comp} shows the period spacings at selected stages,  represented by the hydrogen fraction at the centre, $X_{\rm c}$. The period vs. period-spacing relation has a gentle modulation in early stages of main-sequence evolution, caused by the gradient of hydrogen abundance ($\mu$-gradient) \citep{miglio2008}. The modulation `wavelength' and amplitude decrease as evolution proceeds (see the Appendix for discussions on the modulations). The mean separation, $\Delta P_{\rm g}$, decreases monotonically with evolution (or with decreasing central hydrogen  $X_{\rm c}$).  Because the relation between $\Delta P_{\rm g}$ and $X_{\rm c}$ is monotonic, $\Delta P_{\rm g}$ can be used to constrain the evolution stage; i.e., for a given set of initial parameters we can uniquely choose a model that reproduces the observed value of  $\Delta P_{\rm g}$. The mean period spacing $\Delta P_{\rm g}$ of KIC~9244992 is $0.0264$~d ($2280$~s), which indicates that the hydrogen fraction at the centre  of the star is already reduced to $\sim$0.1. We note that the previous star we studied, KIC~11145123, has $\Delta P_{\rm g} =0.0241$~d ($2082$~s) \citep{kurtz14}, corresponding to the TAMS contraction stage; i.e., the star is slightly more evolved than KIC~9244992. 

\begin{figure}
\centering
\includegraphics[width=0.98\linewidth,angle=0]{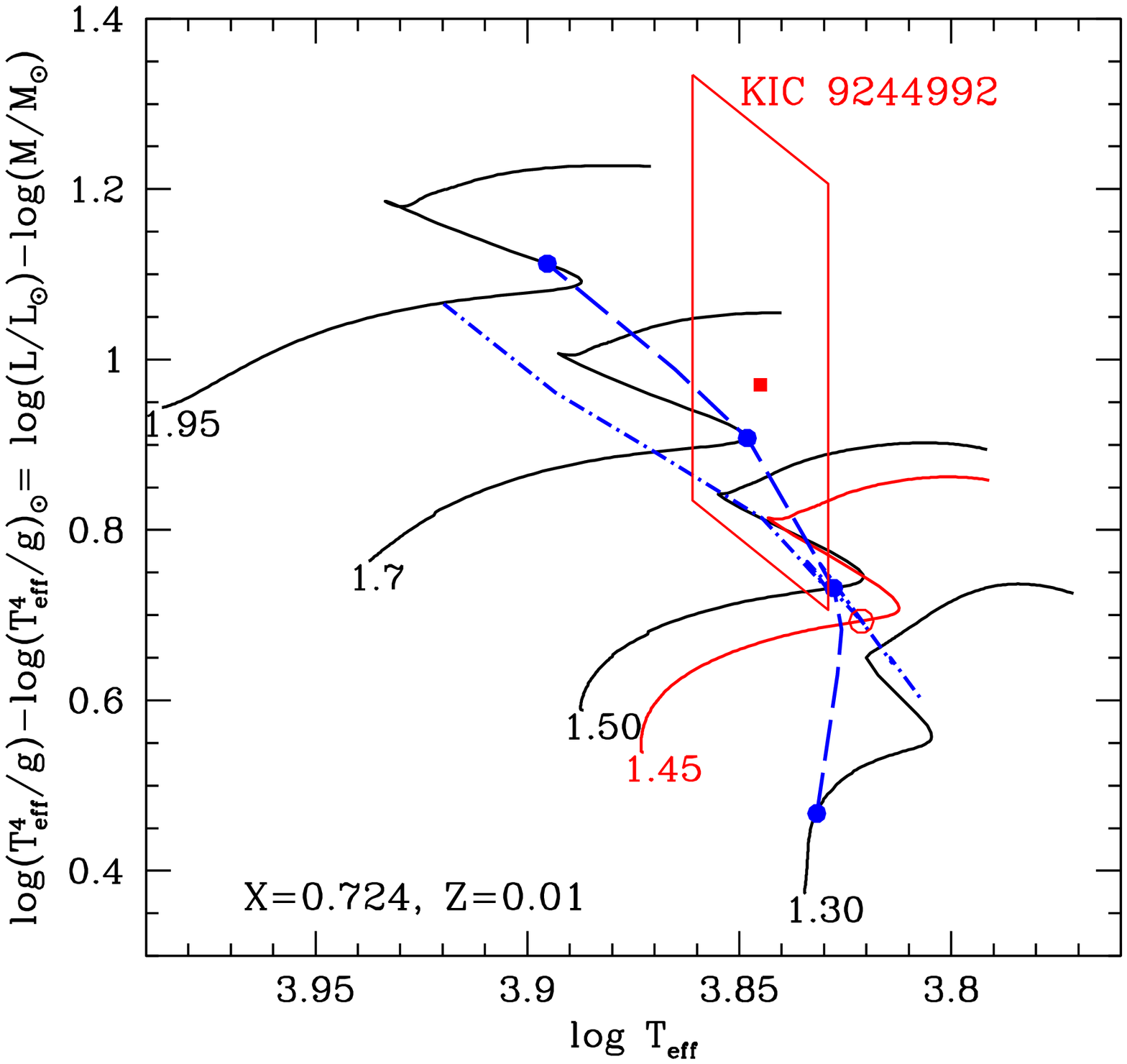} 
\caption{Evolutionary tracks for selected masses with the initial composition $(X,Z)=(0.724,0.01)$ are shown; black lines are without overshooting and the red line is with a small overshooting.  The position of  KIC~9244992 is also plotted. This is a spectroscopic HR (sHR) diagram introduced by \citet{langer14}, where the vertical axis is $\log (L/M)$, equivalent to $4\log T_{\rm eff}-\log g$, evaluated in solar units.  This is convenient when no reliable distance estimates are available. The filled circles (connected with a dashed line) on each evolutionary track indicate the evolutionary stage where the mean period spacing of high-order g~modes is close to the observed value, $0.0264$~d ($2280$~s). The dash-dotted line indicates the locus where the frequency of the radial fundamental mode is equal to the observed highest amplitude frequency $\nu_1$ of KIC~9244992. The open circle indicates the position of our best model for KIC~9244992.
}
\label{fig:shrd}
\end{figure}

Fig.~\ref{fig:shrd} shows evolutionary tracks for some initial masses and the estimated position of KIC~9244992 with an error box, in the spectroscopic HR (sHR) diagram \citep{langer14}, which we use because there are no reliable estimates of the distance of KIC~9244992. Filled circles connected by a dashed line in Fig.~\ref{fig:shrd} show the positions where $\Delta P_{\rm g}\approx 0.0264$~d ($2280$~s).

An additional constraint is that one of the radial modes of the model should fit to the highest amplitude singlet frequency $\nu_1$. The dash-dotted line in Fig.~\ref{fig:shrd} shows the locus on which the frequency of  the fundamental mode is equal to $\nu_1$.
The model at the crossing point between the dash-dotted and dashed lines satisfies the constraints on both $\Delta P_{\rm g}$ and $\nu_1$. This is the best model for a given set of chemical composition $(X,Z)$ and the overshooting parameter $h_{\rm ov}$. Fig.~\ref{fig:shrd} indicates that for the parameter set $(X,Z,h_{\rm ov})=(0.724,0.01,0.0)$, the filled circle on the \mbox{1.5-M$_\odot$} evolutionary track is the best model. However, there may be better models with other parameters, which could be evaluated by comparing with other p~modes.
 
We note that if $\nu_1$ were fitted with the first overtone, the corresponding dash-dotted line would shift rightwards. There would then be no crossing point with the dashed line; this negates the possibility of assigning the first overtone to $\nu_1$.

\subsection{Searching for a best model}

Identifying p~modes in $\delta$~Sct stars is notoriously difficult because of complex frequency distributions. The difficulty is considerably reduced for $\delta$~Sct/$\gamma$~Dor hybrids such as KIC~9244992 and KIC~11145123 by the presence of regularly spaced high-order g~modes, which constrain the evolution stage of the star as discussed in the previous section.  In this subsection, we search for a better model of KIC~9244992 by taking  advantage of the hybrid character.

We searched for a best set of  parameters $(X, Z, h_{\rm ov})$ that reproduces all the observed p~modes. We examined models with parameters combining the three sets $Y = (0.266, 0.30), Z=(0.014,0.010,0.007)$, and $0.0 \le h_{\rm ov} \le 0.01$, where $X=1-Y-Z$. The overshooting parameter $h_{\rm ov}$ determines the scale-length ($h_{\rm ov}H_p$ with $H_p$ being the pressure scale length) for an exponential decay of mixing efficiency \citep{herwig00} above the convective core boundary. 

For each parameter set $(Y, Z, h_{\rm ov})$, there is one model (mass and evolutionary stage) that approximately reproduces both $\nu_1$ and $\Delta P_{\rm g}$~. In order to obtain the most appropriate parameter set, we compare how well these models reproduce the other observed frequencies in the p-mode range, $\nu_2 \ldots \nu_{11}$. We assigned a mode with the closest frequency to each of the observed frequencies, allowing the range of angular degrees $0 \le l \le 6$, although the range is restricted to $l \le 3$ for $\nu_2 \ldots \nu_6$. Then, we calculated  variances defined as: 
\begin{equation} 
{\rm variance} = \sum_{i=2}^{11}\left[\nu_i({\rm obs}) - \nu_i({\rm model})
\over \nu_i({\rm obs})\right]^2 .
\label{eq:variance}
\end{equation}

\begin{figure}
\centering
\includegraphics[width=0.95\linewidth,angle=0]{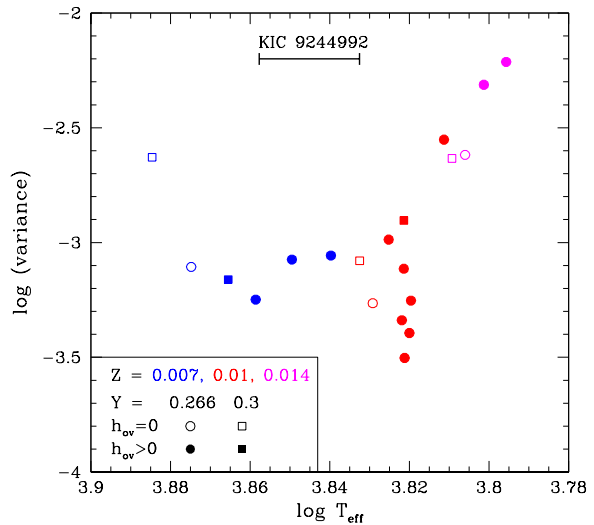} 
\caption{The variances between the observed frequencies and models (which approximately reproduce $\nu_1$ and $\Delta P_{\rm g}$) with various parameter sets, $(Y, Z, h_{\rm ov})$ are plotted with respect to the effective temperatures of the models. The metallicity, $Z$, is colour coded, two values of $Y$ are distinguished by a circle and  a square, and  open symbols are for models without core overshooting. The estimated $T_{\rm eff}$ range of KIC~9244992 is shown by a horizontal line.
}
\label{fig:variance}
\end{figure}

Fig.~\ref{fig:variance} shows variances of models for various parameter sets versus  effective temperatures. Generally, the effective temperature of the models is higher for lower $Z$.  The models with a standard solar metallicity of $Z=0.014$ have variances larger than the other cases, and they are too cool compared with the estimated effective temperature of KIC~9244992. This indicates KIC~9244992 to be metal-poor compared to the Sun, which is consistent  with the KIC data [Fe/H]=$-0.15 \pm 0.3$.

The variance is minimum for a model at $\log T_{\rm eff} = 3.82$ with a parameter set of $(M,Y,Z,h_{\rm ov})=(1.45~{\rm M}_\odot, 0.266, 0.010,0.005)$; detailed parameters are listed in  Table~\ref{table:bestmodel}. The model is slightly metal-poor and evolved to near the end of the main-sequence stage ($X_{\rm c}=0.145$). Small core-overshooting operates in the model with $h_{\rm ov}=0.005$ (which corresponds to an effective mixing range of about $0.05~H_p$; i.e., $\alpha_{\rm ov}\approx 0.05$, as discussed below). The effective temperature of the model, $\log T_{\rm eff}=3.821$, is slightly lower than the estimated range of KIC~9244992 ($\log T_{\rm eff}=3.845\pm 0.012$).  In general,  models having $\log T_{\rm eff} \approx 3.82$ tend to agree with observed frequencies better than the models in other temperature ranges (Fig.~\ref{fig:variance}). The position of the best model in the spectroscopic HR diagram is shown by the red open circle in Fig.~\ref{fig:shrd}. Compared to the position derived from the spectroscopic analysis by  Nemec et al. (in preparation), our best model is lower in $\log L/M$ (vertical axis) and slightly cooler.  

\begin{figure}
\includegraphics[width=0.98\linewidth,angle=0]{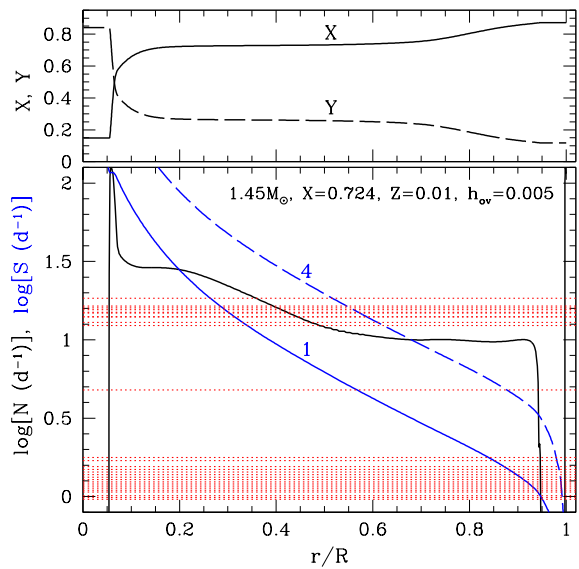} 
\caption{
Upper panel: Distributions of hydrogen ($X$) and helium ($Y$) abundances as a function of the fractional radius, $r/R$, in our best model. Helium is somewhat depleted in outer layers because of atomic diffusion. Lower panel: Radial dependence of the Brunt-V\"ais\"al\"a frquency, $N$ (black line), and the Lamb frequency, $S=\sqrt{l(l+1)}c_{\rm s}/r$ (blue lines), where $c_{\rm s}$ is the sound speed. The number along each blue line indicates the corresponding value of  $l$. Observed frequencies (central frequencies for multiplets) are shown by red dotted lines.
}
\label{fig:prop}
\end{figure}

The propagation diagram of our best model in Fig.~\ref{fig:prop} indicates that all non-radial modes obtained in the p-mode range are influenced by both p-mode and g-mode cavities (i.e., mixed modes), while all g-mode triplets are pure g~modes.

\begin{figure*}
\centering
\includegraphics[width=0.9\linewidth,angle=0]{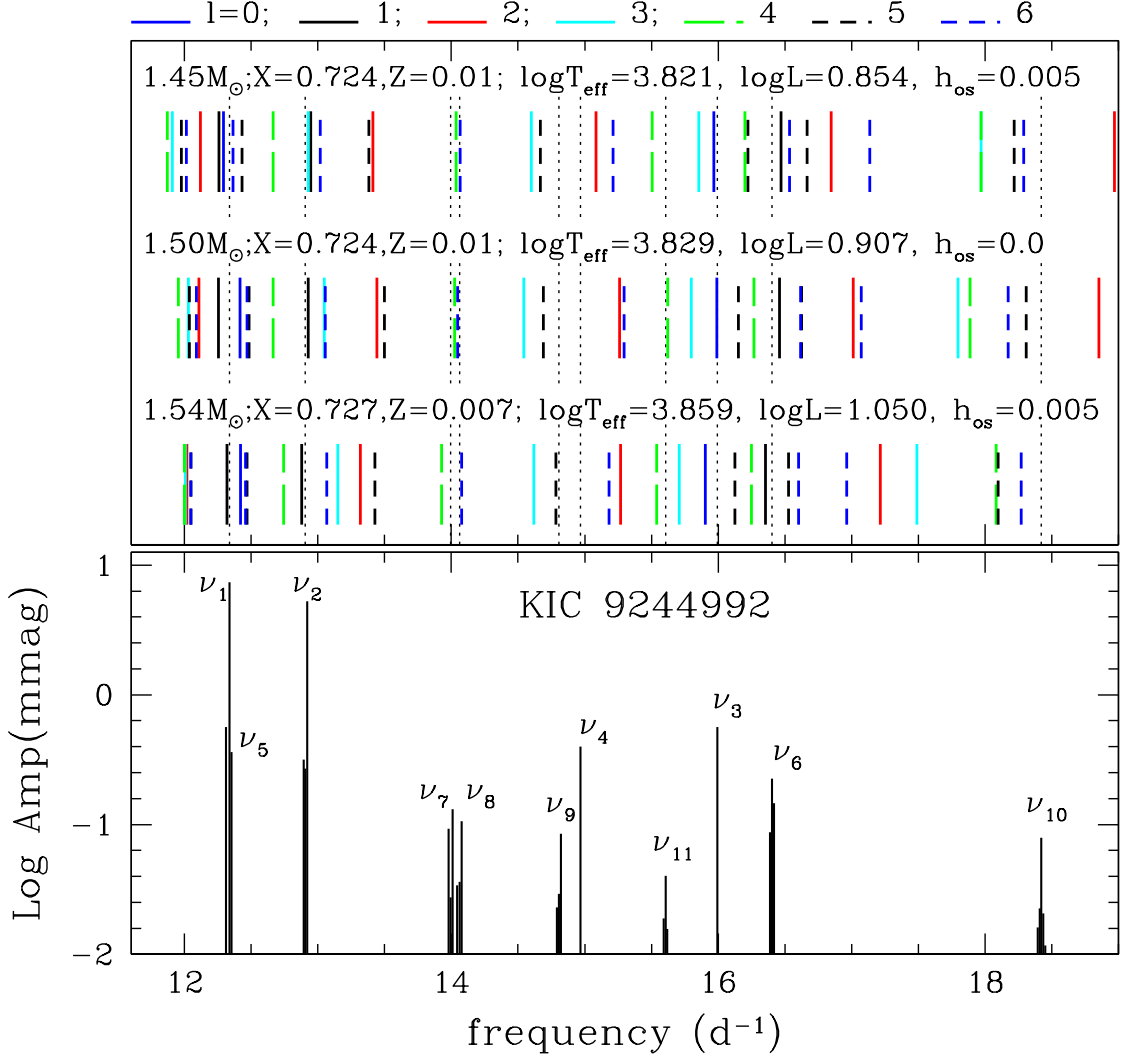} 
\caption{
Observed frequencies in the p-mode range (lower panel) are compared with theoretical ones from the best model (top row in the upper panel) and two other models.  Modes of large latitudinal degree, $l\ge4$, are needed for the lowest amplitude frequencies of $\nu_7 \ldots \nu_{11}$, in all the models examined.  
}
\label{fig:bests}
\end{figure*}

Fig.~\ref{fig:bests} compares observed p-mode frequencies of KIC~9244992 with the best model and two other models. The model shown in the top row in the upper panel of Fig.~\ref{fig:bests} is our best model, whose parameters are given in Table~\ref{table:bestmodel}. The best model reproduces all the detected mode frequencies to better than 1~per~cent, although we have to fit the observed smallest amplitude frequencies of $\nu_7 \ldots \nu_{11}$ with pulsation modes of high latitudinal degrees, $4\le l \le6$. (The same requirement occurs in all the models examined.)

\begin{table}
\centering
\caption[]{ Best model for KIC~9244992}
\begin{tabular}{cccccc}
\hline
$M/{\rm M}_\odot$ & $\log T_{\rm eff}$ & $\log L/{\rm L}_\odot$ &$\log R/{\rm R}_\odot$ & $\log g$   
& Age~(yr)\\
1.45 & 3.821 & 0.854 & 0.308 & 3.982 & $1.9\times10^9$ \\
\hline
$X_{\rm c}^*$ & $X$ & $Y$ & $Z$ & $h_{\rm ov}$ \\
0.145 & 0.724  & 0.266 & 0.010 & 0.005 \\
\hline
\end{tabular}
\leftline{
$^*$ $X_{\rm c}=$ central hydrogen abundance}
\label{table:bestmodel}
\end{table}

Based on our best model, each observed frequency is identified as given in Table~\ref{table:mode_id}, where the radial order $n$ is based on the scheme of \citet{takata2006} (negative and positive $n$ order g~modes  and p~modes, respectively). While $\nu_7$ and $\nu_{11}$ are detected as triplets, we have to fit these with $l=4$ modes because of the absence of  smaller $l$ modes around $\nu_7$ and $\nu_{11}$. The condition is the same for the other three models shown in Fig.~\ref{fig:bests}. 

To fit lowest amplitude frequencies of  $\nu_7$ to $\nu_{11}$ we have to employ modes with high angular degrees  $4 \le l \le 6$. A high $l$ mode is expected to have low visibility because of  cancellation on the stellar surface.  According to \citet{daszynska02}, for a given maximum amplitude on the surface, the visibility of $l=4$ and $l=6$ modes are about 2~per~cent and 0.5~per~cent of a radial mode, respectively. On the other hand, amplitudes of  $\nu_7$ and $\nu_{11}$ are 1.8~per~cent and 0.5~per~cent of  $\nu_1$, comparable to the theoretical estimates for $4\le l \le 6$. This justifies identifying  $\nu_7  \ldots \nu_{11}$ as modes of $l=4$ to $6$.

Two modes are assigned to the triplet $\nu_2$: $(l,n)=(1,-1)$ and $(3,-5)$, both modes have very similar frequencies to each other (Table~\ref{table:mode_id}). The triplet $\nu_2$ has a peculiar character: the amplitude of the main component, which corresponds to the prograde mode ($m=1$), is larger by an order of magnitude than the amplitudes of the other two components (Table~\ref{table:9244992_p}). This peculiar property could possibly indicate that the triplet $\nu_2$  is part of a superposition of  the $(l,n)=(1,-1)$ and $(3,-5)$ modes.

Our best model has a small overshooting of $h_{\rm ov}= 0.005$ from the convective core. The efficiency of mixing in the overshooting zone is assumed to decay exponentially as $\exp[-2z/(h_{\rm ov}H_p)]$ \citep{herwig00} with $z$ being the distance from the boundary of the convective core.
A simpler and more often adopted assumption is to mix fully a zone of thickness $\alpha_{\rm ov}H_p$ above the convective core boundary.
To examine differences in the seismological properties among models adopting different treatments of overshooting, we calculated models with a set of parameters $(M,X,Z) = (1.45~{\rm M}_\odot, 0.724, 0.01)$, the same as that of our best model, but with the simpler assumption for the overshoot mixing controlled by $\alpha_{\rm ov}$. 
We found that the evolutionary track calculated with $\alpha_{\rm ov}=0.052$ is nearly identical to that shown in Fig.~\ref{fig:shrd} (red line) for our best model with $h_{\rm ov}= 0.005$.  We found that a model having very close $L$ and $T_{\rm eff}$ to those of our best model also has very similar pulsation frequencies and hence a  very similar  period/period-spacing relation of g modes.
Therefore, we conclude that our results do not depend on the mixing model of overshooting.

\begin{table*}
\centering 
\caption[]{Mode identification of the observed modes based on our best model. The observed frequencies, $\nu_{\rm obs}$, in the second column are those of the central ($m=0$) component in each multiplet (given in Tables \ref{table:9244992_g}, \ref{table:9244992_g_add} and \ref{table:9244992_p}),
whereas the rotational splittings, $\Delta\nu_{\rm rot}$, in the third column are calculated by $\Delta\nu_{\rm rot}=
\left[
\nu\left(m\right)-\nu\left(-m\right)
\right]/\left(2m\right)$
with $m=2$ for $\nu_{10}$ and $m=1$ for the other modes.}
\begin{tabular}{lrcrcrlc}
\hline
&
\multicolumn{1}{c}{$\nu_{\rm obs}$} &
\multicolumn{1}{c}{$\Delta\nu_{\rm rot}$} &
\multicolumn{1}{c}{$\nu_{\rm mod}$} &
\multicolumn{1}{c}{$l$} &
\multicolumn{1}{c}{$n$} &
\multicolumn{1}{c}{$C_{n,l}$} &
\multicolumn{1}{c}{$\langle P_{\rm rot}\rangle $} \\
&
\multicolumn{1}{c}{d$^{-1}$} &
\multicolumn{1}{c}{$10^{-2}$d$^{-1}$} &
\multicolumn{1}{c}{d$^{-1}$}&
&&&
\multicolumn{1}{c}{d}
\\
\hline
\hline
g & $0.960$ & $0.789\pm 0.012$ & $0.960$ & $1$ & $-38$& $0.497$ & $63.8\pm 1.0$\\
g & $0.985$ & $0.788\pm 0.012$ & $0.986$ & $1$ & $-37$& $0.497$ & $64.1\pm 1.0$ \\
g & $1.011$ & $0.787\pm 0.012$ & $1.012$ & $1$ & $-36$& $0.496$ & $64.0\pm 1.0$ \\
\\
g & $1.068$ & $0.789\pm 0.012$ &$1.070$ & $1$ & $-34$& $0.496$ & $63.9\pm 1.0$ \\ 
g & $1.099$ & $0.787\pm 0.012$ & $1.102$ & $1$ & $-33$& $0.496$ & $64.0\pm 1.0$ \\
g & $1.132$ & $0.791\pm 0.012$ & $1.136$ & $1$ & $-32$& $0.496$ & $63.7\pm 1.0$ \\
g & $1.166$ & $0.790\pm 0.012$ & $1.171$ & $1$ & $-31$& $0.495$ & $63.9\pm 1.0$ \\
g & $1.204$ & $0.791\pm 0.012$ & $1.209$ & $1$ & $-30$& $0.495$  & $63.8\pm 1.0$ \\
g & $1.243$ & $0.792\pm 0.012$ & $1.249$ & $1$ & $-29$ & $0.495$ & $63.8\pm 1.0$ \\
g & $1.285$ & $0.790\pm 0.012$ & $1.292$ & $1$ & $-28$ & $0.495$ & $63.9\pm 1.0$ \\
g & $1.331$  & $0.789\pm 0.012$ & $1.338$ & $1$ & $-27$& $0.495$ & $64.0\pm 1.0$ \\
g & $1.380$  & $0.790\pm 0.012$ & $1.388$ & $1$ & $-26$& $0.494$ & $64.1\pm 1.0$ \\
g & $1.434$  & $0.793\pm 0.012$ & $1.441$ & $1$ & $-25$& $0.494$ & $63.8\pm 1.0$\\
g & $1.491$  & $0.792\pm 0.012$ & $1.499$ & $1$ & $-24$ & $0.494$ & $63.9\pm 1.0$\\
g & $1.554$  & $0.797\pm 0.012$ & $1.562$ & $1$ & $-23$& $0.494$ & $63.5\pm 1.0$ \\
\\
g & $1.697$  & $0.796\pm 0.012$ & $1.705$ & $1$ & $-21$& $0.493$ & $63.7\pm 1.0$ \\
g & $1.777$  & $0.790\pm 0.012$ & $1.786$  & $1$ & $-20$& $0.493$ & $64.2\pm 1.0$\\
\\
g & $4.789$ & $0.791\pm 0.013$ & $4.812$ & $1$ & $-7$ & $0.496$ & $63.7\pm 1.0$\\
\hline
$\nu_1$ &  $12.339$ && $12.299$ & $0$ & $1$
 & 
\multicolumn{1}{c}{---}
 \\
 \hline
$\nu_5$ &  $12.33?$ && $12.254$ & $1$ & $-2$ & $0.229$\\
 \hline
$\nu_2$ &  $12.906$ & $1.377\pm 0.012$ & $12.942$  & $1$ & $-1$ & $0.287$ & $51.8\pm 0.5$\\
 &  & & $12.916$ & $3$ & $-5$ & $0.0390$  & $69.8\pm 0.6$ \\
\hline
$\nu_7$ & $13.996$ & $1.544\pm 0.012$ & $14.052$ & $4$ & $-6$ & $0.0266$ & $63.0\pm 0.5$ \\
\hline
$\nu_8$ & $14.062$ & $1.739\pm 0.012$& $14.062$ & $6$ & $-9$ & $0.0112$ & $56.9\pm 0.4$ \\
\hline
$\nu_9$ & $14.806$ & $1.441\pm 0.013$ & $14.589$ & $3$ & $-4$& $0.0718$ & $64.4\pm 0.6$ \\ 
&&& $14.680$ & $5$ & $-7$ & $0.0150$ & $68.3\pm 0.6$ \\
\hline
$\nu_4$ & $14.966$ && $15.102$ & $2$ & $-2$ & $0.158$ \\
\hline
$\nu_{11}$ & $15.605$ & $1.464\pm 0.014$ & $15.500$ & $4$ &  $-5$ & $0.0511$ & $64.8\pm 0.6$ \\
 \hline
$\nu_3$ & $15.993$ && $15.973$ & $0$ & $2$ & 
\multicolumn{1}{c}{---} \\
 \hline
$\nu_6$ & $16.405$ & $1.504\pm 0.012$ & $16.477$ & $1$ & $1$
 & $0.00653$ & $66.0\pm 0.5$ \\ 
 \hline
$\nu_{10}$ &  $18.423$ & $1.398\pm 0.008$ & $18.218$ & $5$  & $-4$ & $0.0141$ & $70.5\pm 0.4$ \\ 
&&& $18.309$ & $6$ & $-5$ & $0.00695$ & $71.0\pm 0.4$ \\
\hline
\hline
\end{tabular}
\label{table:mode_id}
\end{table*}

\section{Internal rotation}
\label{sec:internal_rotation}

\subsection{The mean rotation period of each mode}

Table~\ref{table:mode_id} lists observed rotational splittings $\Delta \nu_{\rm rot}$ and the Ledoux constants $C_{n,l}$ of the corresponding modes
\citep[see e.g.][for the definition of $C_{n,l}$]{kurtz14}, from which a mean rotation period $\langle P_{\rm rot} \rangle$ for each mode is derived as
\begin{equation} \langle P_{\rm rot} \rangle = {1\over \Delta\nu_{\rm rot}}(1-C_{n,l}) .
\label{eq:Prot}
\end{equation}
While the statistical errors in the frequencies are given in Tables \ref{table:9244992_g}, \ref{table:9244992_g_add} and \ref{table:9244992_p},
the maximum possible errors due to contamination of unresolved peaks can be estimated by 
\begin{equation}
 \epsilon_{T}
=
 \frac{1}{4T} = 1.7 \times 10^{-4}\;
\mbox{d$^{-1}$}
\label{eq:eps_T}
\end{equation}
\citep{kallinger08}, in which $T$ is the total span of the observation period, $1459$\,d.
We also take $\epsilon_{T}$ into account when we calculate the mean rotation period of each mode
based on equation (\ref{eq:Prot}). Note that $\epsilon_{T}$, which is a measure of the frequency resolution of the data set,  is much larger than the statistical errors, which depend on the signal-to-noise ratio. 
For simplicity, we neglect the uncertainty in $C_{n,l}$. The mean rotation period thus obtained is given in the last column of Table~\ref{table:mode_id}.

\begin{figure}
\centering
\includegraphics[width=0.99\linewidth,angle=0,clip]{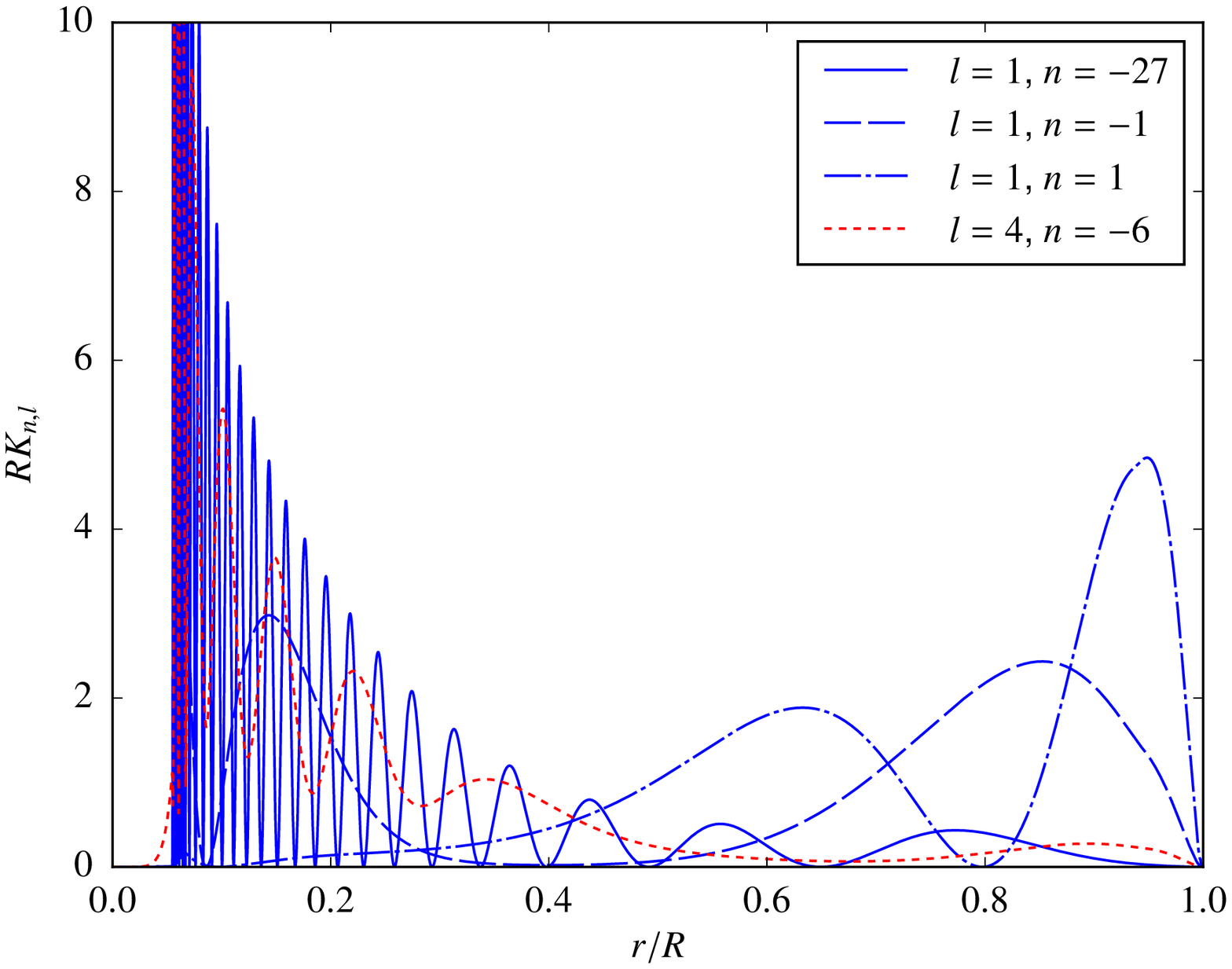}\\ 
\includegraphics[width=0.99\linewidth,angle=0,clip]{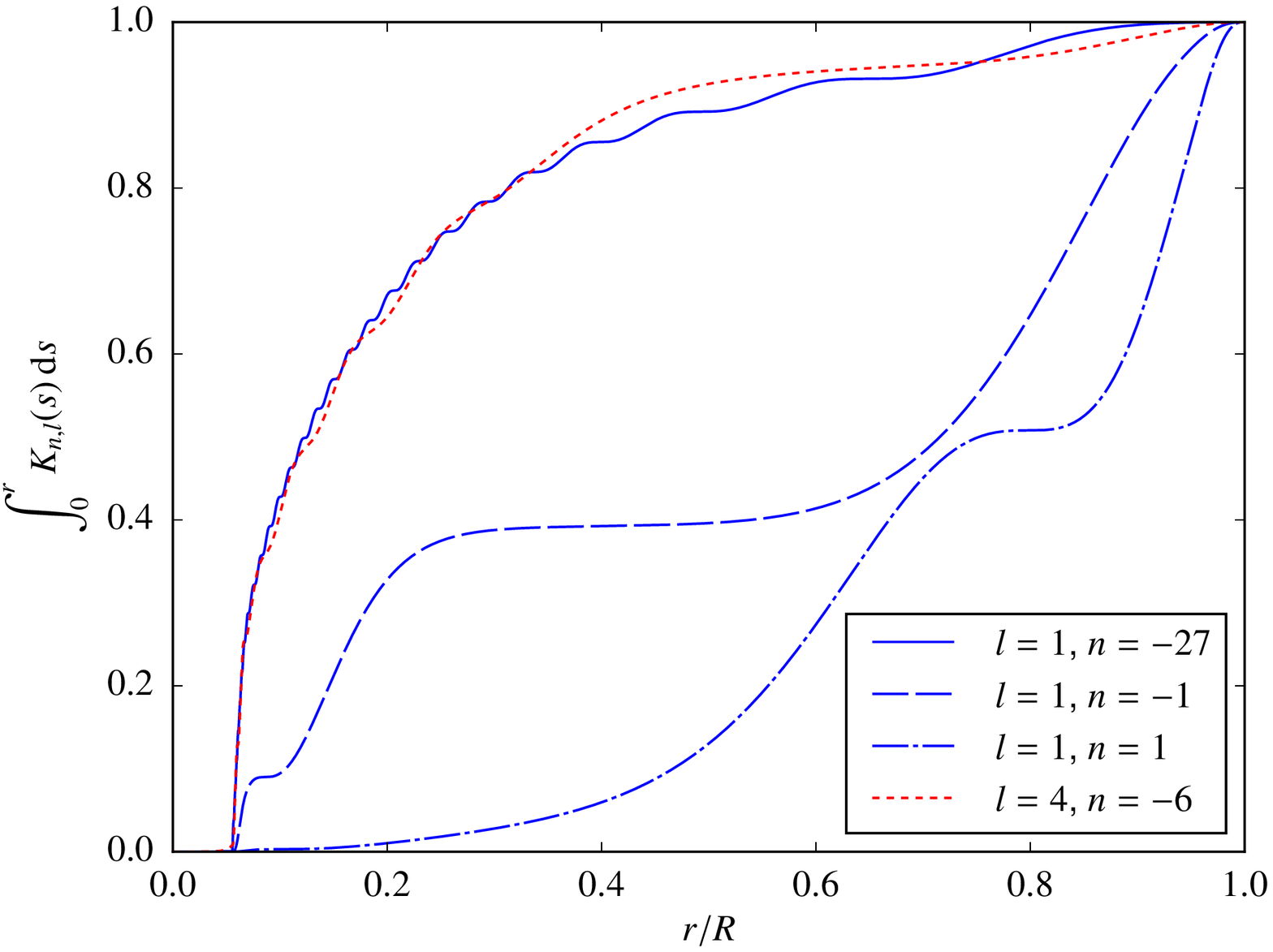} 
\caption{Rotation kernels of selected modes in our best model with 1.45~M$_{\odot}$ as functions of fractional radius.
The upper panel shows the kernels themselves,
whereas the lower panel represents
their cumulative profiles.}
\label{fig:kernel}
\end{figure}

Fig.~\ref{fig:kernel} shows rotational kernels in our best model for a g~mode and three modes identified with the triplets in the p-mode range of KIC~9244992. The average of  the mean rotation period, $\langle P_{\rm rot} \rangle$, obtained from the dipole g~modes is $63.9\pm 0.2$~d.
This represents the rotation period of the core of KIC~9244992.

The frequency $\nu_7$ is identified as the $(l,n)=(4,-6)$ mode. This is a mixed mode that has predominantly g~mode character with no radial node in the p-mode range, although the frequency lies in the mixed-mode range (Fig.~\ref{fig:prop}). The g-mode character of the $l=4$ mode is apparent in the kernel, the cumulative one in particular, which is very similar to that of the dipole g~mode shown in Fig.~\ref{fig:kernel}. The mean rotational period derived for $\nu_7$ is $63.0\pm 0.5$~d (Table~\ref{table:mode_id}), which is consistent with the one obtained from the dipole g~modes within two sigma.

On the other hand, the rotational splitting of the $(l,n)=(1,1)$ mode (assigned to $\nu_6$) represents almost exclusively the rotation rates of outer half (in radius).  The mean rotational period obtained from $\nu_6$ is 
$66.0 \pm 0.5$~d, slightly longer
than that from the dipole g~modes. This indicates that the outer half rotates slightly more slowly than the inner half of the star (by less than 5 per cent).

The mean rotational splitting of $\nu_2$ is $0.0138\pm 0.0001$~d$^{-1}$, slightly smaller than the other triplets in the p-mode range.  
The frequency is identified with the $(l,n)=(1,-1)$ mode, which is a mixed mode having both g-mode and p-mode characters.
The mixed mode character is apparent in the cumulative kernel of the mode in Fig.~\ref{fig:kernel}; about 40~per~cent of the rotational splitting comes from the layers of $r/R\la0.25$ and 60~per~cent from the layers of $r/R \ga 0.6$.  Because of the mixed mode character, the Ledoux constant, $C_{-1,1} = 0.287$ is  much larger than those of typical p~modes. Using this value and the mean rotational splitting of $\nu_2$, we obtain 
$51.8\pm 0.5$~d for a mean rotation period represented by $\nu_2$, which is considerably shorter than the mean rotation periods obtained by the other modes. However, this estimate is rather uncertain for two reasons; 1) the frequency and $C_{n,l}$ of a mixed mode are sensitive to the model structure, and 2) there is another mode $(l,n)=(3,-5)$ whose frequency is very close to $\nu_2$, which gives a rotation period of $69.8\pm 0.6$~d.

\subsection{Two-zone modelling}

\begin{figure}
\centering
\includegraphics[width=0.99\linewidth,angle=0]{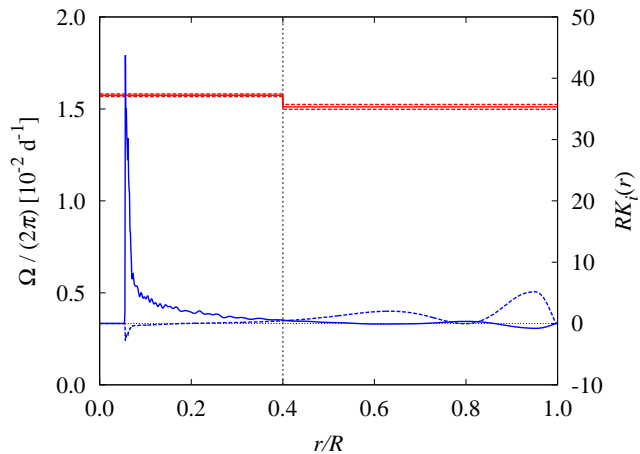} 
\caption{%
The profile of the cyclic rotation frequency, $\Omega/(2\pi)$, inferred by two-zone modelling (red lines) 
and the corresponding averaging kernels multiplied by the total radius, $R$ (blue lines).
The red dashed lines 
designate the (one-sigma) error level.
The blue solid curve represents $K_1\left(r\right)$ for the core rotation,
while the blue dashed curve means $K_2\left(r\right)$ for the envelope rotation.
The boundary between the core and the envelope is set to the fractional radius of $0.4$,
which is indicated by the vertical dotted line.
 }
\label{fig:avkernels}
\end{figure}

We have estimated the rotation profile based on the observed rotational splittings of  the dipole modes except for $\nu_2$ (because of the uncertainties mentioned above). We adopted a two-zone model that assumes constant rotation rates in the core and the envelope. The details of the analysis are described in section 4.3 of \cite{kurtz14}. We set the boundary between the two zones at $r/R = 0.4$, which corresponds to the fractional concentric mass ($M_r/M$) of $0.94$, because the averaging kernels are well-localised in this case. We obtained the rotation period for each zone as
\begin{equation}
P_{\rm rot} =\left\{\begin{array}{ll}
 63.51 \pm 0.28~{\rm d}
  & {\rm for} ~0.0 < r/R < 0.4 \\
 66.18 \pm 0.58~{\rm d}
  & {\rm for} ~0.4 < r/R < 1.0 . \\
\end{array}\right.
\end{equation}
The profile of the rotation rate and the averaging kernels are shown in Fig.~\ref{fig:avkernels}. This result confirms the conclusion in the previous subsection that KIC~9244992 rotates slowly and nearly uniformly, and that the core rotates slightly faster than the envelope. We note that the result has little dependence on the exact location of the boundary of the two zones.
For example, if the boundary is set at $r/R = 0.6$, the rotation periods of
the inner and outer zones are estimated to be $63.63 \pm 0.26$~d and $66.97
\pm 0.79$~d, respectively.

To attempt better localisations of the averaging kernels, we applied the optimally localised averaging method \citep[see, e.g.][for a succinct review in a helioseismic context]{sekii97} to the same mode set. Probably due to the lack of low-order g~modes, however, we were unable to produce significantly better averaging kernels. We therefore prefer the two-zone modelling result for its simplicity.

\section{Discussion and Conclusion}
\label{sec:discussion}
Our best model for KIC~9244992 is slightly metal-poor and has a mass of 1.45~M$_\odot$. The star could be an SX~Phe star as suggested by \citet{balona12}. If we assume that [Fe/H]$\approx \log (Z/Z_\odot)$ and adopt $Z_\odot = 0.014$, then $Z=0.010$ corresponds to [Fe/H]$= -0.15$, which agrees with the value in the KIC, although this argument is not strong because of the uncertainty on the KIC value of $\pm 0.3$. However, the metal deficiency is not sufficient to identify the star as an SX Phe star; \citet{mcnamara2011} suggests [Fe/H]$\lesssim -1$ for SX Phe stars. Hence, it might be appropriate to regard KIC~9244992 as a $\delta$~Sct/$\gamma$~Dor star having a slightly metal-poor composition.  

In order to examine the excitation/damping of pulsations in F stars, and to make predictions for the instability range of $\gamma$ Dor stars in the HR diagram, the effect of the interaction between convection and pulsation must be  taken into account. \citet{dupret2005} performed such analyses and found that the excitation of g~modes in  models with mixing length twice the pressure scale height ($\alpha =2$) is consistent with  the observed range of $\gamma$~Dor variables \citep[although that distribution is still controversial:][]{grigahcene2010,uytterhoeven2011,tkachenko2013}. Our best model is slightly cooler than the theoretical red edge for $\alpha=2$, although the effective temperature spectroscopically estimated for KIC~9244992 is consistent with the theoretical instability range.  The $1.6$-M$_\odot$ model of \citet{dupret2005} predicts the excitation of dipole g~modes with periods longer than about $0.7$~d (the upper limit depends on $T_{\rm eff}$), while, in KIC~9244992,  almost consecutive dipole g~modes between periods of $0.56$~d  and $1.04$~d and a dipole mode at $0.2$~d are excited,  hence a few short-period dipole modes detected in KIC~9244992 are outside the predicted period range. The theoretical model also predicts the excitation of quadrupole ($l=2$) g~modes with periods longer than about $0.5$~d. This is consistent with the identification of two outliers at periods of $0.66$~d and $0.87$~d in Fig.~\ref{fig:k9244992echelle} as quadrupole g~modes ($n=-42$ and $n \sim -54 \sim -57$, respectively). We also note that there are still many low-amplitude peaks remained in the g-mode range (bottom panel of Fig.~\ref{fig:9244992_ftg}), that could be $l\ge2$ g~modes.

The rotational splittings of p~modes indicate that the surface of KIC~9244992 rotates slowly, with a period of $\approx 66$~d or longer.
Combining the rotation period and the radius $\approx2~R_\odot$ of our best model yields an equatorial rotation velocity of  $1.5$~km~s$^{-1}$.  It will be challenging to measure such a small rotational velocity by spectroscopic analysis. The slow and nearly uniform rotation found for KIC~9244992 is similar to the main-sequence star KIC~11145123 \citep{kurtz14}, whose equatorial rotation velocity is also found to be $\sim 1$~km~s$^{-1}$. 
Both stars are near the end of main-sequence stage.

In the wholly convective Hayashi phase, stars probably rotate rigidly because of large turbulent viscosity. 
After the Hayashi phase, if  angular momentum is conserved locally in radiative layers, the central region rotates faster than the envelope, because of  an increase in the central mass concentration. The difference would become about factor 3 at the ZAMS, and about factor 10 near the end of the main-sequence evolution \citep{kurtz14}. Clearly, a strong mechanism for angular momentum transport must be acting during the main-sequence stage to result in the nearly rigid rotation observed in KIC~9244992 and KIC~11145123. The results obtained for these two stars may indicate that low-mass stars, in general, may rotate slowly and nearly rigidly at the very end of the main-sequence evolution. This proposition must be confirmed (or negated) by asteroseismic analyses for many $\delta$ Sct/$\gamma$ Dor hybrid stars in the future. If true, it has important implications not only for the angular-momentum transport during main-sequence evolution, but also for the internal rotation of sub-giants and red giants, for which A-F stars are progenitors.
 
It was found recently in several sub-giants and red giants that the contrast of rotation rates between the core and the envelope is  much weaker than theoretically expected \citep{deheuvels2012, deheuvels2014}. Angular momentum must be redistributed much faster than the present models predict, during evolution before stars become red giants. Stimulated by the discovery, some theoretical attempts to explain this observational fact have been made. For example, \citet{eggenberger2012} pointed out the necessity of an additional mechanism of angular momentum transport during main sequence and post-main sequence evolution in order to reproduce the weak contrast of rotation between the core and the envelope.  
\citet{marques2013} modeled angular momentum transport by parametrizing turbulent diffusivity and computed a series of stellar evolutionary models. 
They found it hard to reproduce the observed property of weak contrast of rotation between the core and the envelope, and concluded that much more efficient mechanisms of angular momentum transport should be at work to explain rotation periods of red giant cores. A similar conclusion was reached  by \citet{Cantiello2014} after detailed calculations taking account of rotationally induced instabilities and circulations, as well as magnetic fields.
The role of internal gravity waves has also been reviewed by \citet{Mathis2013} and \citet{Fuller2014}.  

We have found in this paper that KIC~9244992, an F star near the TAMS, rotates slowly and nearly rigidly. The result is similar to that obtained by \citet{kurtz14} for KIC~11145123.  If this property is general; i.e., if  all A-F stars ($1.3 \lesssim M/M_{\odot} \lesssim 2$) rotate uniformly at the end of main sequence, angular-momentum transport during the main-sequence evolution must be much more efficient than included in the present evolution models. In addition, evolutionary models to reproduce slow core rotation in giants also has to reproduce the uniform rotation at the TAMS. 
It is important to analyse many more $\delta$ Sct/$\gamma$ hybrid stars, as well as to search for efficient mechanism of angular momentum transport.

\section*{acknowledgements}
We thank NASA and the {\it Kepler} team for their revolutionary data. This work was initiated with support from a JSPS Japan-UK Joint Research grant. 
D. W. Kurtz thanks the JSPS for a Furusato Award that partially funded this work. We thank Bill Paxton and the MESA project team for developing the efficient stellar evolution code MESA. H.~Saio and M.~Takata were partially supported by JSPS KAKENHI Grant Number 26400219.

\bibliography{k9244992c}

\appendix
\section{Period spacings of g~modes}

The period spacings of high-order g~modes are nearly uniform because the period $P_n$ is given as
\begin{equation}
P_n \approx {2\pi^2 |n|\over\sqrt{l(l+1)}}\left[\int_a^b{N\over r}dr\right]^{-1}
\end{equation}
\citep[e.g.,][]{unno1989,aerts2010}, where $N$ is the Brunt-V\"ais\"al\"a frequency, and $a$ and $b$ are the radii at the lower and the upper boundary of the propagation zone of the g~mode. As the evolution proceeds, the core of a star contracts and hence the radius $a$ decreases, which increases the integral of $N/r$ and in turn decreases  periods and period spacings of g~modes. A buildup of a $\mu$-gradient zone ($\mu=$ mean molecular weight) enhances the evolutionary change of $P_n$. In addition, \citet{miglio2008} found that the $\mu$-gradient zone produces a modulation in the period and period-spacing relation and that the `wavelength' of the modulation decreases with evolution. They also found the amplitude of the modulation decreases significantly if turbulent diffusive mixing is included.

\begin{figure}
\centering
\includegraphics[width=0.90\linewidth,angle=0]{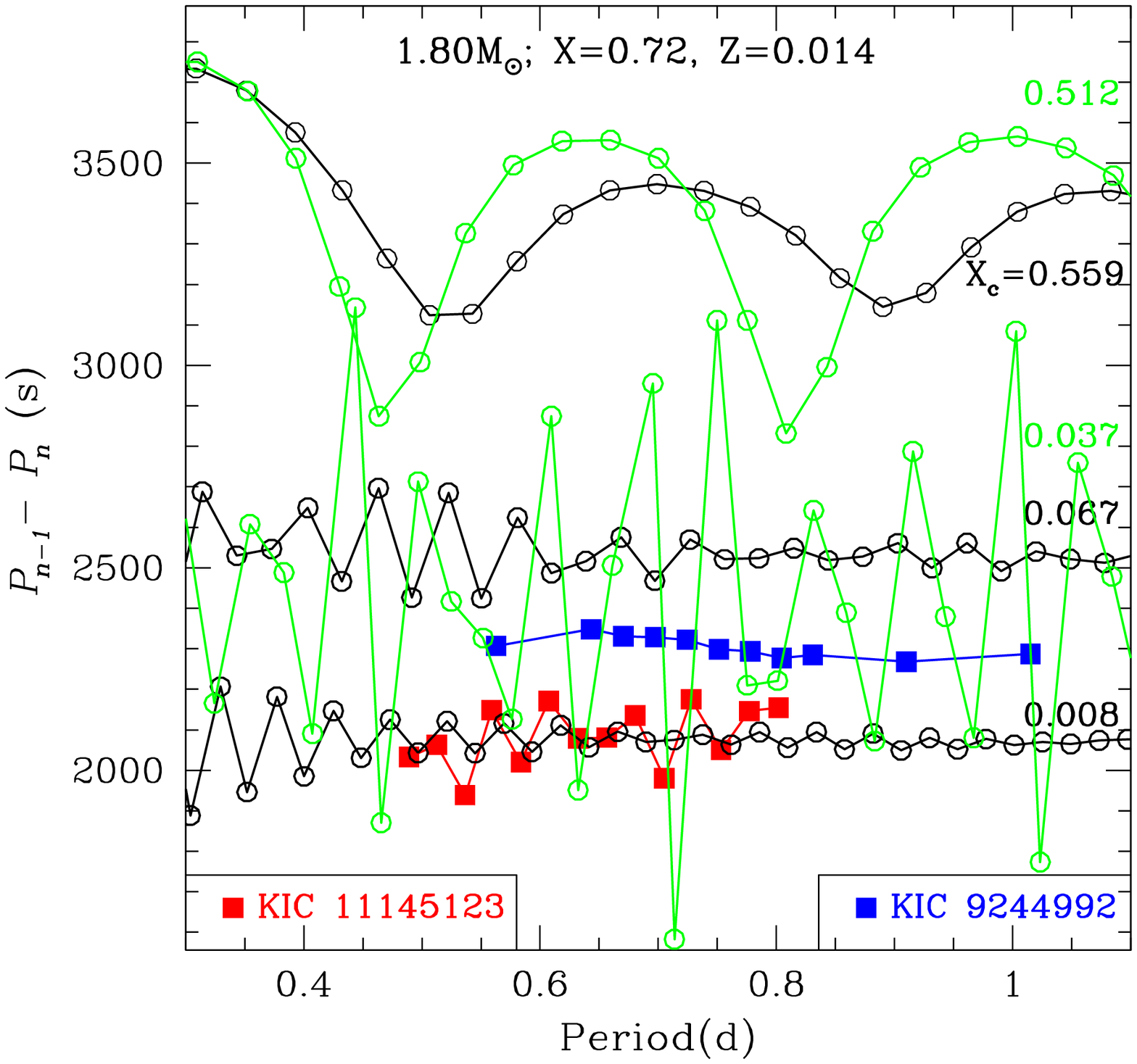} 
\caption{Period vs. period-spacing relations at selected evolutionary stages are compared between models with (black lines) and without (green lines) atomic diffusion. Also shown are the relations of KIC~9244992 and KIC~11145123. The number along each line indicate the central hydrogen abundance. }
\label{fig:pdp_dif}
\end{figure}

\begin{figure}
\centering
\includegraphics[width=0.90\linewidth,angle=0]{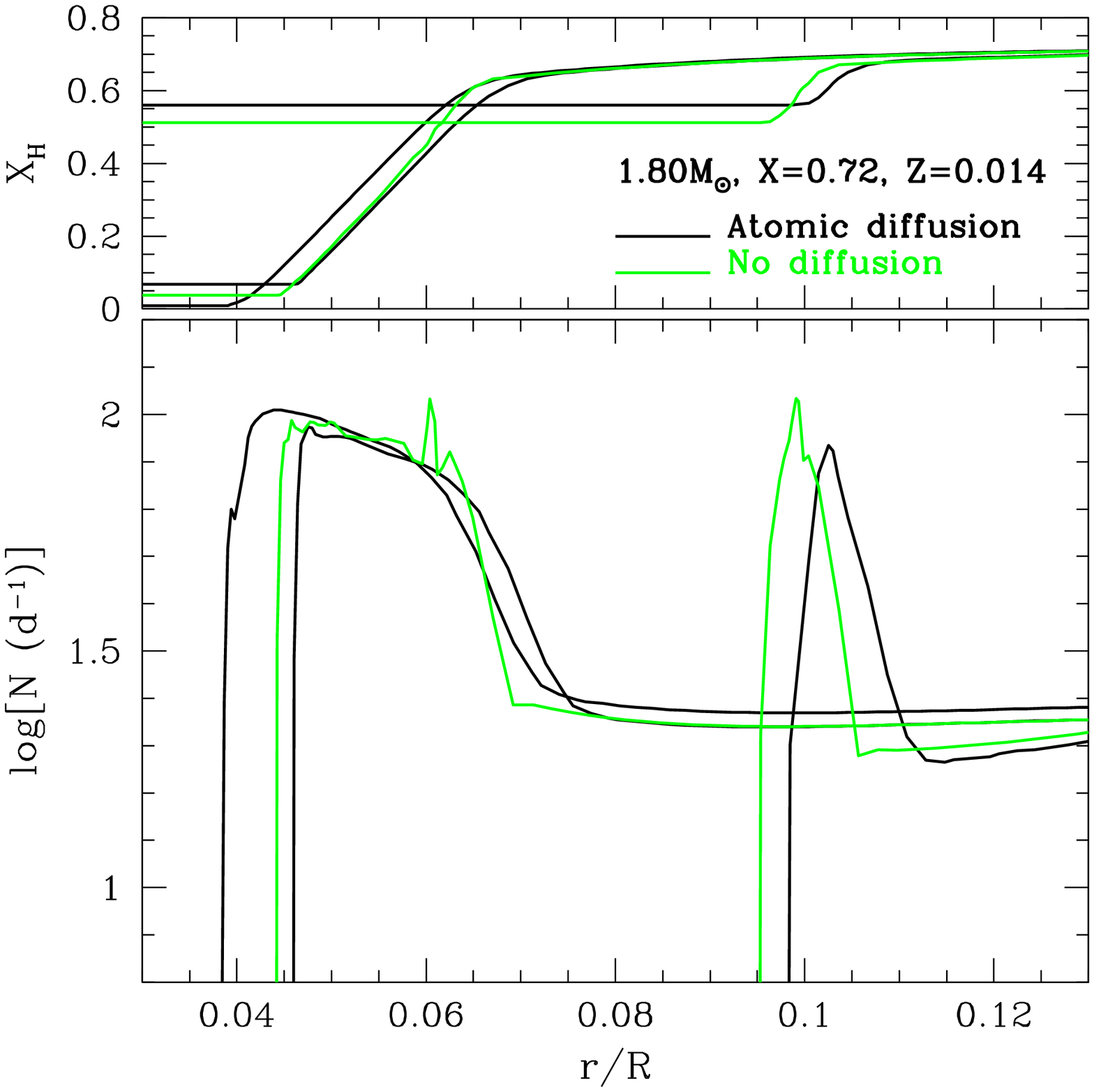} 
\caption{The runs of hydrogen abundance (upper panel) and Brunt-V\"ais\"al\"a frequency (lower panel) at the stage corresponding to each curve in Fig.~\ref{fig:pdp_dif}.}
\label{fig:prop_X}
\end{figure}

In this appendix we briefly discuss the effect of atomic diffusion, which we have included in our evolution calculations to obtain a smooth distribution of  the Brunt-V\"ais\"al\"a frequency. Fig.~\ref{fig:pdp_dif} shows period spacings of g~modes as a function of the period for selected models; black and green lines are for models with and without atomic diffusion, respectively. Fig.~\ref{fig:prop_X} shows the runs of hydrogen abundance (upper panel) and of the Brunt-V\"ais\"al\"a frequency for each model shown in Fig.~\ref{fig:pdp_dif}. In the late stage of main-sequence evolution, the period spacing varies wildly as a function of the period if the atomic diffusion is not activated; this is in stark contrast to the cases with atomic diffusion. The difference is caused by a moderate difference in the gradient of the run of  the Brunt-V\"ais\"al\"a  frequency at the outer boundary of the $\mu$-gradient zone (wiggles in $N$ are not important); smoothing of the distribution by atomic diffusion results in a large reduction of the oscillations of the period spacings, which is consistent with the observed relations of KIC~9244992 and KIC~11145123. This indicates that in a real star some mechanism of smoothing the distribution of chemical abundance must be at work -- atomic diffusion or turbulent diffusive mixing. In our two slowly rotating stars, atomic diffusion seems more appropriate. 

\end{document}